\documentclass[final, a4paper, 10pt]{article}
\usepackage{natbib}
\usepackage{amsmath}
\usepackage{amsthm}
\usepackage{amsfonts}
\usepackage{latexsym}
\usepackage{graphicx}
\usepackage{epstopdf}
\DeclareMathAlphabet{\mathpzc}{OT1}{pzc}{m}{it}
\renewcommand{\Re}{\mathbb{R}}

\author{%
\\
\small
}
\date{ }
\title{Basic statistics\\ for distributional symbolic variables: \\ a new metric-based approach}
\begin{document}
\maketitle

\begin{abstract}
In data mining it is usual to describe a group of measurements using summary statistics or through  their empirical distribution functions. Each summary of a group of measurements is the representation of a typology of individuals (sub-populations) or of the evolution of the observed variable for each individual. Therefore, typologies or individuals are expressible through multi-valued descriptions (intervals, frequency distributions). Symbolic Data Analysis, a relatively new statistical approach, aims at the treatment of such kinds of data.\\
In the conceptual framework of Symbolic Data Analysis, the paper aims at presenting new basic statistics for numeric multi-valued data. First of all, we propose how to consider all numerical multi-valued descriptions as special cases of distributional data, i.e. as data described by distributions. Secondly, we extend some classic univariate (mean, variance, standard deviation) and bivariate (covariance and correlation) basic statistics taking into account the nature, the source and the interpretation of the variability of such data. As opposed to those proposed in the literature, the novel statistics are based on a distance between distributions, the $\ell_2$ Wasserstein distance.\\
Using a clinic dataset, we compare the proposed approach to the existing one showing the main differences in terms of interpretation of results.
\end{abstract}

\section{Introduction}\label{sec:intro}
In many real experiences data are collected or represented by multi-valued descriptions: intervals, frequency distributions, histograms, density distributions, and so on. Typical examples are the description of macrodata in official statistics, basic statistics, frequency distributions or estimate of parameters from data referred to a group of units or to the same unit observed in multiple occasions, or data directly measured in a condition of uncertainty. In these cases, even if we are observing a single variable, the information coming from the observations couldn't be conveniently expressed by only one number or category. Thus, it is usual to represent such information by multiple values, and such multi-valued description is the start of novel analysis.
Several proposals appeared in the literature for processing such data, according to their nature, source or mathematical modeling. For example, when data are expressed by intervals of $\Re$, Interval arithmetic \citep{Moore1996}, fuzzy set \citep{Moore2003}, or a Symbolic Data Analysis \citep{BoDi2000} approach provide useful tools for their statistical treatment. When the data domain is categorical, a noteworthy approach is the Compositional data one \citep{ait86}. Among the methods listed above, Symbolic Data Analysis (SDA) \cite{BoDi2000,BilDid2006, bildid03, NoiBri2012} approach provides models and techniques for generalizing the statistical treatment of most of them. In fact, SDA is a relatively new statistical approach designed for processing data described by set-valued variables (or {\em Symbolic variables}) like interval, multi-valued discrete, multi-categorical, histogram and modal variables. In particular, modal variables can model the description of an individual, of a group, or of a concept, by distribution of probabilities, frequencies or, in general, by random variables.\\
In recent years, several authors proposed and defined new statistics and new techniques for the analysis of a particular case of modal data description: the histogram-valued data.
\citet{BerGoup2000} proposed a first set of basic univariate and bivariate statistics that was integrated and extended by \citet{BilDid2006}. Further developments for the quantification of the variability and the dependence between variables of a set of multi-valued data can be found in \cite{Billard2007} and \cite{Brito2007}. For interval data, the statistics  proposed by \citet{Billard2007} and \citet{BerGoup2000} start from the assumption that an interval-valued data $[a, b]$ is a uniformly distribution $U\sim (a,b)$. Considering histograms as a weighted collection of intervals, \citet{BerGoup2000} and \citet{BilDid2006} extended the univariate (mean, variance and standard deviation) and bivariate statistics (covariance and correlation) to histogram data.\\
A set of multi-valued data holds two kind of variability: an internal to data variability and a between data variability. The first is related to the multiplicity of values that describes the single observation: for example, an interval description has proper variability related to its width. The last is related to the different multi-valued descriptions: two intervals can be different for position, width or both. The basic univariate statistics proposed by \citet{BerGoup2000} and \citet{BilDid2006} are not sensible to express the role of the two sources of variability (for example, the variance of a set of identical multi-valued data is in general positive). In this paper,  we propose a novel set of univariate and bivariate statistics that better take into account the two sources of variability and extend some properties of the classic basic statistics to those for multi-valued numeric data.\\

The paper is organized as follows: in section \ref{sec:data} we present the different multi-valued numerical data according to their definition in SDA, and how to consider them as a unique, more general, type of data described by distributions. In section \ref{sec:Univariate_stats}, we show state-of-the-art basic univariate statistics and their coincidence with the basic statistics of finite mixture of distributions, we reflect on their use and we propose new basic statistics that solve some discrepancies present in the former approach. The novel univariate statistics emerge from the definition of a measure of variability that is related to a distance between distributions. Among the different distances presented in the literature, we motivate the choice of the $\ell_2$ Wasserstein distance \cite{rusch_2001}  showing the gain of interpretability of the results provided by this choice and its consistency with the double source of variability of a set of multi-valued data.\\
The choice of the $\ell_2$ Wasserstein distance allows to use a novel product operator between two distributions. Using such an operator, in section \ref{sec:interdep}, we propose an extension of the classical covariance and correlation measures between two standard variables to the case of numeric modal variables. Also in this case, we show that it is possible to take into account the different source of variability of the multi-valued data.\\
Using a clinic dataset presented in \cite{BilDid2006}, in section \ref{sec:apply} we present an application of the proposed statistics and a comparison with those proposed in the same book. The results give evidence of the interpretative properties of the novel statistics. Section \ref{sec:concl} ends the paper with some comments and suggestions for future research.

\section{Numerical symbolic modal data}\label{sec:data}
The definition of the types of data used in this paper is presented consistently with the Symbolic Data Analysis (SDA)\citep{BoDi2000,BilDid2006} terminology.
SDA aims to extend classical data analysis and statistical methods to more complex data called \emph{symbolic data} that are realizations of a so-called \emph{symbolic variables}. In SDA the \emph{symbolic datum} describes an individual according to a set of numbers or categories, that can be equipped with a set of weights, while \emph{standard datum} describe an individual assigning a single measurement (number or category) of a standard variable. \citet{BoDi2000} defined \emph{symbolic variables} as follows:
\newtheorem{rmk}{Definition}
\begin{rmk}
Let $E$ be a set of objects, a variable $Y$ is termed \emph{set-valued} with domain $\mathcal{Y}$, if for all $i\in E$,\begin{equation}\label{eq:setvalued}
    \begin{array}{c}
       Y:E\rightarrow D \\
       i\mapsto y(i)
     \end{array}
\end{equation}
where the description $D$ is defined by $D=\mathpzc{P}(\mathcal{Y})=\{U\neq \emptyset|U\subseteq \mathcal{Y}\}$.
A set-valued variable $Y$ is called \emph{multi-valued} if its description set $D_c$ is the set of all finite subsets of the underlying domain $\mathcal{Y}$; such that $|y(i)|< \infty$, for all $i\in E$.

A set-valued variable $Y$ is called \emph{categorical multi-valued} if it has a finite set $\mathcal{Y}$ of categories and \emph{quantitative multi-valued} if the values $y(i)$ are finite sets of real numbers.

A set-valued variable $Y$ is called \emph{interval-valued} if its description set $D_I$ is the set of intervals of $\Re$.
\end{rmk}
\citet{BoDi2000} also defined \emph{modal (symbolic) variables} as follows:
\begin{rmk}\label{def_mod}
A modal variable $Y$ on a set $E$ of objects with domain $\mathcal{Y}$ is a mapping
\begin{equation}\label{modal1}
    y(i)=(S(i),\pi_i),\forall i\in E
\end{equation}
where $\pi_i$ is a measure or a (frequency, probability or weight) distribution on the domain $\mathcal{Y}$ of possible observation values (completed by a $\sigma$-field), and $S(i)\subseteq \mathcal{Y}$ is the support of $\pi_i$ in the domain $\mathcal{Y}$. The description set of a modal variable is denoted with $D_m$.
\end{rmk}

In the present paper, it is not considered the \emph{multi-categorical} case, but only those descriptions based on numerical support.
We propose to treat all numerical (single-valued or set-valued) variables as particular cases of the modal variables. In particular, we propose to treat data in a probabilistic perspective, as \emph{distributional data}.
In order to follow the terminology adopted in SDA, the variables which allow distributions as description of individuals are termed \emph{modal-numeric} (probabilistic) variables.
\begin{rmk}
Given a set $E$ of objects with domain $\mathcal{Y}$ and support $S(i)$ partitioned into $n_i$ subsets, a probability measure associated with a density function $\psi$ and with the respective distribution function $\Psi$, such that
\begin{equation}\label{wheight}
    \Psi_i\left({y=S_h(i)}\right)=\int_{S_h(i)}{\psi_i(z)dz}
\end{equation}
where $h=1,...,n_i$, a modal (probabilistic) variable $Y$ is a mapping
\begin{equation}\label{modal2}
    y(i)=\left(\; {S_h\left(i\right),\Psi_i\left(S_h\left(i\right)\right)}\; \right),\forall i\in E.
\end{equation}
\end{rmk}

In the following, we consider the main types of symbolic numeric variables. After defining the support $S(i)$, the density function $\psi$ and the distribution function $\Psi$, we propose how to consider them as particular modal-numeric descriptor.
\begin{description}
  \item[Classic single valued data] $S(i)=y_i$ such that $y_i\in \Re$, and $\pi_i=1$

  In this case, the individual $i\in E$ is described by a single value $y_i$. $y(i)=y_i$ is considered like a modal-numeric datum associated with a density function that follows as Dirac delta function shifted in $y_i$:
 \[
\psi_i(y) =\delta(y-y_i) =\left\{ {\begin{array}{lr}
   {+\infty}&{if\hspace{0.3cm}y=y_i}  \\
   {0} & {otherwise}\\
\end{array}} \right.
\]
     subject to the constraint that $\int_{-\infty}^{+\infty}{\delta(z-y_i)dz}=1$.

     The corresponding distribution function is:
     $$
     \Psi_i(y)=   {\int_{-\infty}^{y}{\delta(z-y_i)dz}}
     $$

     Therefore the modal-numeric description is:
     $$
     y(i)=(y_i,\Psi_i(y_i))=(y_i,1).
     $$
     \item[Multi-valued discrete description] Modal multi-valued discrete description can be considered as a mixture of Dirac delta distributions, where $S(i)$ is a set of distinct single values.

       The support can be written as  $S(i)  = \left\{ {y_{1i} ,...,y_{li} ,...,y_{n_ii} } \right\}$ where, each element of the support is associated with a $\pi_{li}$, such that $\sum\limits_{l=1}^{n_i}{\pi_{li}}=1$ (or the mixing weights). We then consider the function:
$$
\psi_i(y)=\sum\limits_{l=1}^{n_i}{\pi_{li}\delta(y-y_{li})}
$$
where $\psi _{i}(y)$ is a density function associated to the
description of $i$ and the corresponding distribution function is:
$$
\Psi_i(y)=\sum\limits_{l=1}^{n_i}\left({\pi_{li}\int_{-\infty}^y {\delta(z-y_{li})dz}}\right).
$$
In this case, the modal-numeric description is:
     $$
     y(i)=\{(y_{1i},\pi_{1i}),\ldots,(y_{n_i i},\pi_{n_i i})\}.
     $$
  \item[Interval description] $S(i)=[a_i,b_i]$ such that $a_i\leq y_i \leq b_i$, and assuming a uniform distribution in $S(i)=[a_i,b_i]$, we can rewrite $\pi_i$ as
   \[
\psi_i(y) = \left\{ {\begin{array}{lr}
   {{\frac{1}{b_i-a_i}}}&{if\hspace{0.3cm}a_i\leq y \leq b_i}  \\
   {0} & {otherwise}\\
\end{array}} \right.
\]
The corresponding distribution function is:
\[
\Psi_i(y) = \left\{ {\begin{array}{lr}
{0}&{if\hspace{0.3cm}y < a_i}  \\
   {\int_{a_i}^y{\frac{1}{b_i-a_i}dz}}&{if\hspace{0.3cm}a_i\leq y \leq b_i}  \\
   {1} & {if\hspace{0.3cm}y > b_i}\\
\end{array}} \right.
\]
     In this case, the modal-numeric description is:
     $$
     y(i)=([a_i,b_i],\Psi_i(a_i\leq y\leq b_i))=([a_i,b_i],1).
     $$

       If it is known the distribution of the data on the interval we may consider $\Psi_i(y)$ as a the (cumulative) distribution function corresponding to $\psi_i(y)$.
  \item[Histogram valued description] We assume that
$S(i)=[\underline{y}_i;\overline{y}_i]$ (the support is bounded in $\Re$).
The support is partitioned into a set of $n_i$ intervals $S(i)  = \left\{ {I_{1i}
,\ldots,I_{ui} ,\ldots,I_{n_ii} } \right\}$, where  $ I_{li}  = \left[
{\underline{y}_{li} ,\overline{y}_{li} } \right) $ and $l=1,\ldots,n_i$, i.e.
$$
\begin{array}{l}
i. \hspace{10pt}  I_{li}  \cap I_{mi}  = \emptyset ;\hspace{5pt}l \ne m \hspace{5pt}; \\
ii. \hspace{10pt} \bigcup\limits_{l = 1,...,n_i} {I_{li} }  = S(i) \\
 \end{array}
$$
 Histograms suppose that each interval is uniformly dense.
It is possible to define the modal description of $i$ as follows:
$$
y(i) = \{ (I_{li} ,\pi _{li})\;|\;\forall I_{li} \in S(i) ;\;\pi
_{li}=\Psi_i(\underline{y}_{li}\leq y \leq \overline{y}_{li})  = \int\limits_{I_{li}}  {\psi _{i} (z)dz}  \ge 0\}
$$
 where $\int\limits_{S(i)} {\psi _i (z)dz} =1$.\\
Given the generical interval $I_{li}  = \left[
{\underline{y}_{li} ,\overline{y}_{li} } \right]$ where $\underline{y}_{li}<\overline{y}_{li}$, and $U(y|I_{li})=U(y|\underline{y}_{li},\overline{y}_{li})$ as the Uniform continuous function defined between $\underline{y}_{li}$ and $\overline{y}_{li}$, we may rewrite a histogram as a linear combination of Uniform distribution (a mixture) as follows:
$$
\psi_i(y)=\sum\limits_{l=1}^{n_i}{\pi_{li}U(y|I_{li})}
$$
where $\psi _{i}(y)$ is a density function associated to the
description of $i$ and the corresponding distribution function is:
$$
\Psi_i(y)=\sum\limits_{l=1}^{n_i}\left({\pi_{li}\int_{-\infty}^y {U(z|I_{li})dz}}\right).
$$
 In this case, the modal-numeric description is:
     $$
     y(i)=\{(I_{1i},\pi_{1i}),\ldots,(I_{n_i i},\pi_{n_i i})\}.
     $$

  \item[Continuous random variable] $S(i)$ correspond to the support of the random variable, $\psi_i(y)$ correspond to its density function.

       We can consider, then, the density as $$\psi_i(y)=f_i(y|\mathbf{\Theta}),$$ where $\mathbf{\Theta}$ is a vector of parameters, and the distribution function as
       $$ \Psi_i(y)=\int_{-\infty}^y{f_i(z|\mathbf{\Theta})dz}.$$
       In this case the modal-numeric description is:
     $$
     y(i)=(y,\psi_i(y)).
     $$
\end{description}
In conclusion, the numeric \emph{set-valued} variables (single-valued and interval-valued) are considered \emph{distributional} variables (or numeric modal symbolic variables) whose distribution function is a uniform or a $\delta$-Dirac distribution. While the first assumption is accepted in the SDA literature \cite{BerGoup2000}, the last one corresponds to the same assumption for a thin interval: a point-value can be considered a zero-width interval.\\
The proposed reformulation of the  different types of numeric symbolic variables into a unique and more general type of \emph{distributional symbolic variable} (a numerical modal probabilistic symbolic variable)  permits to consider a unique approach for computing univariate and bivariate statistics for a wide class of symbolic numeric data. The rest of the paper discusses the proposal of new statistics for \emph{distributional variables}.

\section{Basic univariate statistics for numerical symbolic data}\label{sec:Univariate_stats}
The first to propose a set of univariate and bivariate statistics for \emph{symbolic data} was \citet{BerGoup2000}, and subsequently \citet{BilDid2006} improved them. The \citet{BerGoup2000} approach  relies on the so-called \emph{two level paradigm} presented in SDA in \cite{BoDi2000}: the set-valued description of a statistical unit of a \emph{higher} order is the generalization of the values observed for a class of the \emph{lower} order units. For example, the income distribution of a \emph{nation} (the higher order unit) is the empirical distribution of the incomes of each citizen (the lower order units) of that nation. Naturally, other generalization of grouping criteria can be taken into consideration.\\
The generalization process from lower to higher order units considered by \citet{BerGoup2000} and by \citet{BilDid2006} implies the following assumptions: given two symbolic data $y(1)$ and $y(2)$ described by the frequency distributions $f_1(y)$ and $f_2(y)$, a lower order unit can be described by a single value $y_0$  that has a probability of occurring equal to $\frac{f_1(y_0)+f_2(y_0)}{2}$. The univariate statistics proposed by \citet{BerGoup2000} and by \citet{BilDid2006} for a symbolic variable (namely, a variable describing higher order units, or a class of units) correspond to those of the classic variable used for describing the (unknown) lower order units. Thus, given a set $E$ of $n$ higher order units described by the numerical symbolic variable $Y$, the mean, the variance and the standard deviation proposed by \citet{BerGoup2000} and extended by \citet{BilDid2006} correspond to those of a finite mixture of $n$ density (or frequency) functions with mixing weights equal to $\frac{1}{n}$. Given $n$ density functions denoted with $\phi_i(y)$ with the respective means $\mu_i=E(Y_i)$ and variance $\sigma^2_i=E[(Y_i-\mu_i)^2]$, and given the finite mixture density $\phi(y)$ as follows:
\begin{equation}\label{mixture}
  \phi(y)=\sum\limits_{i=1}^{n}{\frac{1}{n}\phi_i(y)}={\frac{1}{n}\sum\limits_{i=1}^{n}\phi_i(y)},
\end{equation}
\citet{Uhwir2006} shows that the mean $\mu=E(Y)$ and the variance $\sigma^2=E[(Y-\mu)^2]$ of $\phi(y)$ are the following:
\begin{eqnarray}
  \mu=E(Y)&=&\frac{1}{n}\sum\limits_{i=1}^{n}\mu_i; \label{mu_mix} \\
  \sigma^2=E[(Y-\mu)^2]&=&\frac{1}{n}\sum\limits_{i=1}^{n}{\left(\mu_i^2+\sigma_i^2\right)}-\mu^2. \label{s2_mix}
\end{eqnarray}
It is worth noting that the two statistics in eq. (\ref{mu_mix}) and (\ref{s2_mix}) are the same as those proposed by \citet{BilDid2006} for a numeric symbolic variable, except for a different notation.\\ For the sake of simplicity, we show only the formulas related to interval-valued data. Let $Y$ be an interval-valued variable, thus, the generic symbolic datum is $y(i)=[a_i;b_i]$ with $a_i \leq b_i$ belonging to $\Re$. According to \cite{BerGoup2000}, $y(i)$ is considered as a uniform distribution in $[a_i;b_i]$, with mean equal to  $\mu_i=\frac{a_i+b_i}{2}$ and variance equal to $\sigma^2_i=\frac{\left(b_i-a_i\right)^2}{12}$. Given a set of $n$ units described by a interval-valued variable, the {\em symbolic sample mean} $\bar{Y}$ \cite[eq. (3.22)]{BilDid2006} is:
\begin{equation}\label{mu_Bil}
\bar{Y}=\frac{1}{2\cdot n}\sum\limits_{i=1}^{n}{\left(b_i+a_i\right)}.
\end{equation}
It is straightforward to show its equivalence with $\mu$ in eq.(\ref{mu_mix}), indeed:
$$
\bar{Y}=\frac{1}{n}\sum\limits_{i=1}^{n}\frac{\left(b_i+a_i\right)}{2}=\frac{1}{ n}\sum\limits_{i=1}^{n}\mu_i=\mu .
$$
In \citep[eq. (3.22)]{BilDid2006} is also proposed the {\em symbolic sample variance} as follows:
\begin{equation}\label{var_Bi}
  S^2=\underbrace{\frac{1}{3\cdot n}\sum\limits_{i=1}^{n}\left(b^2_i+b_i\cdot a_i+a_i^2\right)}_{(I)}-\underbrace{\frac{1}{4\cdot n^2}\left[\sum\limits_{i=1}^{n}\left(a_i-b_i\right)\right]^2}_{(II)}.
\end{equation}
Considering that:
\begin{eqnarray*}
\mu _i^2 + \sigma _i^2 &=& {\left( {\frac{{{b_i} + {a_i}}}{2}} \right)^2} + \frac{{{{({b_i} - {a_i})}^2}}}{{12}} = \frac{{{{({b_i} + {a_i})}^2}}}{4} + \frac{{{{({b_i} - {a_i})}^2}}}{{12}} = \\
 &=& \frac{{3b_i^2 + 3a_i^2 + 6{b_i}{a_i} + b_i^2 + a_i^2 - 2{b_i}{a_i}}}{{12}} = \\
 &=& \frac{{4b_i^2 + 4a_i^2 + 4{b_i}{a_i}}}{{12}} = \frac{{b_i^2 + {b_i}\cdot{a_i} + a_i^2}}{3}
\end{eqnarray*}
the term (I) of eq. (\ref{var_Bi}) can be expressed as follows:
$$
\frac{1}{3\cdot n}\sum\limits_{i=1}^{n}\left(b^2_i+b_i\cdot a_i+a_i^2\right)=\frac{1}{n}\sum\limits_{i=1}^{n}\left(\mu_i^2+\sigma_i^2\right) .
$$
The term  $(II)$ is clearly $\mu^2$, indeed:
$$
\frac{1}{4\cdot n^2}\left[\sum\limits_{i=1}^{n}\left(a_i-b_i\right)\right]^2=\left[\frac{1}{n}\sum\limits_{i=1}^{n}\frac{\left(a_i-b_i\right)}{2}\right]^2=\left[\frac{1}{n}\sum\limits_{i=1}^{n}\mu_i\right]^2=\mu^2 .
$$
Thus, $S^2$ in eq. (\ref{var_Bi}) corresponds to eq. (\ref{s2_mix}), indeed:
\begin{equation}\label{var_MIX}
S^2=(I)-(II)=\frac{1}{n}\sum\limits_{i=1}^{n}\left(\mu_i^2+\sigma_i^2\right)-\mu^2=\sigma^2 .
\end{equation}
The same correspondences also hold for the mean and the variance of the other numerical modal symbolic variables.\\
This approach is particularly useful and coherent when the symbolic data, referred to higher order units, are the description of groups of lower level units (the income distribution of a nation is described by the incomes of its citizens). In general, all the symbolic data have the same weight, but knowing in advance the cardinality of the groups, it is possible to estimate an unbiased mean or standard deviation of the variable describing all the lower order units (the per-capita income in Europe is the weighted, by the respective population, mean of the per-capita incomes of the single nations). \\
In some situations, the proposed approach for the definition of the univariate basic statistics hides some peculiarities present in the data. For example, describing the pulse rate of a patient during a particular activity (while he walks, runs, swims, sleeps, etc.), we can modelize this information using the distribution of the pulse rates recorded during that activity. If we collect the same information from $n$ people, we obtain $n$ symbolic data described respectively by $n$ pulse rate distributions.
Using the mean and the standard deviation of a symbolic variable like those proposed by \citet{BilDid2006}, we obtain measures related to all the pulse rate measurements independently from belonging to a particular individual of the group. Indeed, being the basic statistics of a mixture, the pulse rate measurements can be permuted among the individuals and the basic statistics do not change, avoiding the possibility of comparing individuals. In such a case, we could be interested in studying the variability of the individuals according to their pulse rate distributions, such that, the more the pulse rate distributions are different, the more variability is in the data. Extending the concept of variability like a measure of divergence of the observed data with respect to an average datum, the mean individual should have a distribution that is as close as possible to all the observed distribution: the average should be expressed by a distribution. In this case, like for the classic case,  if the $n$ data are identical (thus identical to the mean) the variability of that symbolic variable should be zero. Other dissimilarities for interval-valued data, treated as distributions, have been considered in \citep{IrVe2008}.\\
Therefore, in SDA the source of variability of a symbolic variable is twofold:
\begin{description}
  \item[internal to data] each symbolic datum has an inherent variability due to the summarization process of lower order units into higher order ones, and that is a possible element of the domain of the symbolic variable: each individual is described by the distribution of the recorded pulse rate;
  \item[between data] a set of $n$ units described by a symbolic variable is a set of $n$ multi-valued observations: each individual may have a different pulse rate distribution.
\end{description}
In contrast to \citep{BilDid2006}, where the sample variance of a symbolic variable is the amount of the internal to data and between data variability, we here consider the possibility of relating the internal variability as a characteristic that pertains to the mean unit (the mean individual is described by a distribution that is as close as possible to all the observed pulse rate distributions) while the variability of a symbolic data is related to the diversity of their symbolic descriptions.

\subsection{The mean and the variability of a set of data described by distributions}\label{sec:mean}
While in probability theory the mean corresponds to the expected value of a random variable, in descriptive statistics the mean can assume several definitions. Starting from proximity relations among  data it is possible to define the so called {\em Fr\'{e}chet} means, while starting from the definition of a function of the observed data it is possible to define the so called {\em Chisini} means. More formally:
\begin{description}
      \item[Fr\'{e}chet (or Karcher) mean] according to \cite{Ginest2012}, given  a set of $n$ elements described by the variable $Y$, a $d$ distance between two descriptions and a set $W=(w_1,\ldots,w_n)$ of $n$ real numbers, a {Fr\'{e}chet type} mean (barycenter) $M$ is the argmin of the following minimization problem:
          \begin{equation}\label{eq:FREM}
          M=\arg\min_{x}\sum\limits_{i=1}^n{w_i d^2(y_i,x)}\end{equation} provided that a unique minimizer exists.
      \item[Chisini mean] according to \cite{Chisi1929}, given a set of $n$ units described by the single real valued variable $Y$  and a function $F$, a {\em Chisini type} mean $M$ must satisfy the following condition:
          \begin{equation}\label{eq:CHIM} F(y_1,\ldots,y_i,\ldots,y_n)=F(\underbrace{M,\ldots,M}_{n\;times})\end{equation} for example, the arithmetic mean is invariant with respect to the sum function, i.e.:
          $$\sum\limits_{i=1}^n {y_i}=\sum\limits_{i=1}^n {M}=nM\;  \Rightarrow\; M=\frac{1}{n}\sum\limits_{i=1}^n {y_i}.$$
To extend {\em Chisini type} means to multi-valued numeric data, it is important to define functions and operators for multi-valued data.
\end{description}
The definition of a {\em Fr\'{e}chet} and {\em Chisini} compatible mean of distributional variables requires two conditions: the definition of a distance between distributions (or random variables) and the definition of, at least, the sum of distributions and the product of a distribution and a scalar.

\begin{figure}
  \centering
  \includegraphics[width=\textwidth]{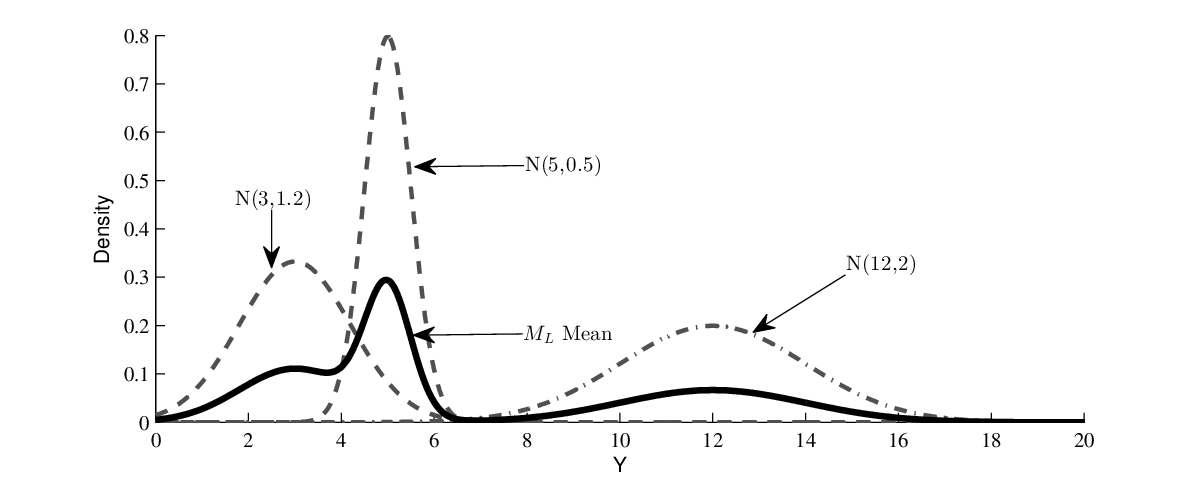}\\
  \caption{The mean according to $L_2$ distance}\label{fig:mixture}
\end{figure}

A variety of dissimilarities for symbolic data are presented in \citep[Chap. 8]{BoDi2000}. For continuous and multi-valued categorical data, several component-wise dissimilarities are presented: the Gowda-Diday, Ichino-Yaguchi and De Carvalho dissimilarities. Unfortunately, none of those are formulated for comparing data described by frequency or probability distribution functions with a numeric and continuous support. In the same chapter, for comparing multi-valued modal data, the authors presented a review of dissimilarities based on particular families of divergence indices for probability distributions.     Such divergences are based on a function of the likelihood ratio between two probability measures and, therefore, they are not symmetric. The well known Kullback-Liebler (KL) divergence suffers from this inconvenience, too. A well-known symmetric version of the KL divergence is the Jensen-Shannon (JS) dissimilarity, which corresponds to the mean of the KL divergences between two probability distributions and their mixture. \citet{Niel_2009}, in a  {\em k-means} framework, studied the  minimization of the sum of squared divergences based on information theory (KL, JS, and their generalizations into Bregman divergences). They showed that in minimizing a distance criterion, a single {\em centroid} distribution of a set of probability distributions cannot be obtained. They solved this problem by proposing a couple of centroids (a left and a right centroid for each class) according to the direction of the computed divergence.\\
  Starting from the study of \citet{GISU02}, \citet{VeIrp2007} considered a set of dissimilarity and distance measures for probability distributions. They observed that not all the considered probabilistic distances and dissimilarities consent to identify a unique distribution as a {\em center} of a set of distributions, or that the resulting {\em center} could not be expressed as a distribution. However, the authors noticed that only two distances give the possibility of defining a single center in the form of a distribution: the $L_2$ Euclidean and the $\ell_2$ Wasserstein distance between distributions.\\
  In the rest of the paper, we use the following notation:
given $n$ probability distributions with density (or probability) functions denoted with $\phi_i(y)$, the respective expected value is denoted with $\mu_i$ and the  standard deviation is denoted with $\sigma_i$; each $\phi_i(y)$ is in a one-to-one correspondence with a cumulative distribution function ({\em cdf}) denoted with $\Phi_i(y)$ and with a quantile function ({\em qf}) denoted with $\Phi^{-1}_i(t)$.\\
\paragraph{The sample mean based on Euclidean distance.} The $L_2$ distance between two density functions  $\phi_i(y)$ and $\phi_{i'}(y)$ (for continuous distributions) is:
\begin{equation}\label{distL2_density}
  d_{L}(\phi_i(y), \phi_{i'}(y))=\sqrt{\int\limits_{-\infty}^{+\infty}\left|\phi_i(y)-\phi_{i'}(y)\right|^2dy}
\end{equation}
it is straightforward to prove that the {\em Fr\'{e}chet} mean associated to $d_{L}$ (assuming equal weights $w_i$) is given by the finite mixture of the $n$ density functions as follows:
\begin{equation}\label{mix_center}
M_L(y)=\arg\min_{x}\sum\limits_{i=1}^n{ d_{L}^2(y(i),x)}=\frac{1}{n}\sum\limits_{i=1}^{n}\phi_i(y)
\end{equation}
and, thus, the mean and the variance of $M_L(y)$ correspond to those presented above in eqs. (\ref{mu_Bil}) and (\ref{var_Bi}). $M_L(y)$  is a density function, but in general has a different shape with respect to the set of summarized densities: for example, a mixture of not identical Normal distributions is not a Normal distribution too. As we can see in Fig. \ref{fig:mixture}, the $M_L(y)$ representation is coherent with the aggregation criterion where each distribution comes from a sub-population and $M_L(y)$ is the distribution of $Y$ for the whole population.\\
\citet{VeIrp2007} considered another distance: the $\ell_2$ version of the $\ell_p$ Wasserstein distance. The literature provides different formulations of the Wasserstein distance but we use the formalization used by \citet{rusch_2001} (which also contains the main references to the Wasserstein metric) which expressed the distance using the quantile functions associated with the respective \emph{cdf}s as follows:
\begin{equation}\label{distW2_density}
  d_{W_p}(\phi_i(y), \phi_{i'}(y))=\left(\int\limits_{0}^{1}\left|\Phi_i^{-1}(t)-\Phi_{i'}^{-1}(t)\right|^p dt\right)^\frac{1}{p}.
\end{equation}
The proposed formulation in \cite{rusch_2001} shows that $d_{W_p}$ distance can be considered as an extension of the classic $L_p$ Minkowski distance for quantile functions (the inverse of {\em cdf}s). The quantile functions ({\em qf}s) have several useful statistical properties \cite{Gilch2000}, some of the most interesting for the arguments of this paper are: {\em qf}s are in a one-to-one correspondence with the corresponding density functions, {\em qf}s have a finite domain ($t\in[0;1]$), and {\em qf}s are non-decreasing functions.

\begin{figure}
  \centering
  \includegraphics[width=\textwidth]{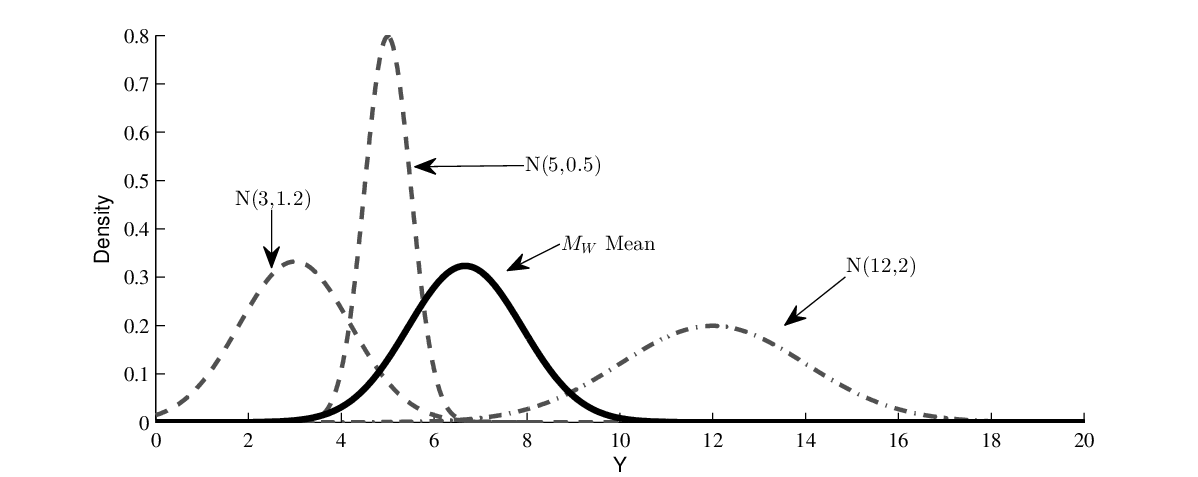}\\
  \caption{The mean according to $\ell_2$ Wasserstein distance}\label{fig:meanWdist}
\end{figure}

\paragraph{The sample mean based on Wasserstein distance.}
For avoiding multiple indices, we denote with $d_W$ the following formulation of the $\ell_2$ Wasserstein distance between two probability distributions:
\begin{equation}\label{eq:WDIS}
    d_W(\phi_i(y) ,\phi_{i'}(y) ) =  \left\{\int\limits_0^1 {\left[ {\Phi_i^{ - 1} (t) - \Phi_{i'}^{ - 1}
(t)} \right]^2dt}\right\}^{1/2}.
\end{equation}
In this case, the {\em Fr\'{e}chet} mean with respect to $d_{W}$ (assuming equal weights $w_i$) is the distribution corresponding to the \emph{mean quantile function} that solves the following optimization problem:
\begin{equation}\label{Wass_prob}
M_W(y)=\arg\min_{x(y)}\sum\limits_{i=1}^n{ d_{W}^2(\phi_i(y),x(y))}.
\end{equation}
Assuming that $x(y)$ is a density function and $\chi^{-1}(t)$ is the corresponding quantile function and considering the integral operator, the solution of the optimization in eq. \ref{Wass_prob} is obtained for each $t\in [0,1]$ according to the classic first order condition, as follows:
\begin{equation}\label{Solution}
\frac{\delta \left[{\sum\limits_{i=1}^n\left( {\Phi_i^{ - 1} (t) - \chi^{-1}(t)} \right)^2}\right]}{\delta \chi^{-1}(t)}=0\; \Rightarrow
\chi^{-1}(t)=\bar{\Phi}^{-1}(t)=\frac{1}{n}\sum\limits_{i=1}^n\Phi_i^{-1}(t),
\end{equation}
where $\bar{\Phi}^{-1}(t)$ indicates the \emph{mean quantile function} observed in $t$.\\
The {\em Fr\'{e}chet} mean distribution corresponds to the distribution that is into a one-to-one correspondence with $\bar{\Phi}^{-1}(t)$, i.e.
\begin{equation}\label{MW_frech}
M_W(y)=\bar{\phi}(y)=\frac{d(\bar{\Phi}^{-1}(t))^{-1}}{dy}=\frac{d\bar{\Phi}(y)}{dy}.
\end{equation}
Figure \ref{fig:meanWdist} shows $M_W(y)$ for the same three Normal distributions represented in Fig. \ref{fig:mixture}. Differently from $M_L(y)$, we observe that $M_W(y)$ has a central position and an intermediate shape with respect to the observed distributional data.
For showing the \emph{centrality} properties of $M_W$, the quantity $\rho _{i,i'}$, the correlation coefficient between two quantile functions, plays an important role. It is defined as follows:
\begin{equation}\label{eq:rhoQQ}
\rho _{i,i'} =  \frac{\int\limits_0^1 {\left( \Phi_i^{ - 1}(t) - \mu _i  \right)\left( {\Phi_{i'}^{ - 1}(t) - \mu _{i'} }
\right)dt} }  {\sigma_i \sigma_{i'} }=\frac{\int\limits_0^1 { \Phi_i^{ - 1}(t) \Phi_{i'}^{ - 1}(t) dt}- \mu _i\mu _j  }  {\sigma_i \sigma_{i'} }.
\end{equation}
It is worth noting that $\rho _{i,i'}$ is always positive, being the correlation coefficient between two not decreasing functions and is exactly equal to 0 when at least one distribution has no variability (is a single valued data). Imaging a QQ (Quantile-Quantile) plot, $\rho _{i,i'}$ is the correlation of the scattered points, therefore it can be considered a measure of the similarity  of the {\em shapes} of two distribution functions. In fact, $\rho _{i,i'}=1$ only if the two distributions have the same standardized quantiles by the respective mean and standard deviation, which occurs when the two distributions have the same shape.\\
 According to \citet{bar99} and using eq. \ref{eq:rhoQQ}, it is possible to prove (see \ref{Appendix}) that the squared $\ell_2$ Wasserstein distance can be decomposed as follows:
\begin{equation}\label{decomp}
d_W^2 (\phi_i(y) ,\phi_{i'}(y) )= \underbrace {\left( {\mu _i - \mu _{i'} }
\right)^2}_{Location} + \underbrace{\underbrace {\left( {\sigma _i - \sigma _{i'}
} \right)^2}_{Size} + \underbrace {2 \sigma_i \sigma_{i'} (1 -
\rho_{i,i'} )}_{Shape}}_{Variability}. \end{equation}
The decomposition permits the interpretation of the (squared) distance between two distribution functions according to two {\em additive} aspects. The {\em Location} aspect emphasizes the difference in position of the  two distributions through the (squared Euclidean) distance between the respective means. The second  aspect is related to the different {\em Variability} structure of the compared distributions due to the different standard deviations (the {\em Size} component) and to the different shapes of the density functions (the {\em Shape} component). While the {\em Size} component is expressed by the (squared Euclidean) distance between the standard deviations, the {\em Shape} component is fundamentally governed by the value of $\rho_{i,i'}$.\\
The decomposition in Eq. (\ref{decomp}) suggests that optimization problem in Eq. (\ref{Wass_prob}) leads to a solution where $M_W$ has the minimum \emph{Location} difference with respect to all the locations of the distributions and the minimum \emph{Variability} difference with respect to all the variabilities of the distributions. Given the optimization problem in Eq. (\ref{Wass_prob}), and considering that the quantities in Eq. (\ref{decomp}) cannot be negative, $\mu_x$ and $\sigma_x$ respectively the mean and the standard deviation of $M_W(y)$, and $\rho(i,x)$ the correlation between the \emph{qf} of $i$-th observation and the  \emph{qf} of  $M_W(y)$ it is possible to show that:
\begin{eqnarray*}\label{Wass_prob}
\mu_x&=&\arg\min_{x}\left[\sum\limits_{i=1}^n\left( {\mu _i - x }\right)^2\right]\\
\{\sigma_x, [\rho_{1,x},\ldots,\rho_{n,x}]\}&=&
 \arg\min_{s,[r_{1,x},\ldots,r_{n,x}]\in [0,1]^n}\sum\limits_{i=1}^n \left[\left( {\sigma _i - s
} \right)^2 + 2 \sigma_i \cdot s \cdot(1 -r_{i,x} )\right].
\end{eqnarray*}

In order to show that $M_W$ is also a {\em Chisini} mean we introduce the sum operator between \emph{qf}s and the product of a \emph{qf} by a scalar. Let $\mathcal{F}$ be the set of functions of the kind $y=f(t)$ with bounded domain $[0;1]$ and imagine in $\Re$. Let $\mathcal{Q}\subset \mathcal{F}$ be the set of all possible  quantile functions, i.e. the set containing only non-decreasing functions with bounded domain in $[0;1]$ . Let  $(f+g)(t)=f(t)+g(t)$ be the sum between two elements of $\mathcal{F}$  and $(k\cdot f)(t)=k\cdot f(t)$ the product of a scalar $k$ by a function, it is known that $(\mathcal{F},+,\cdot)$ is a vector space. However, given the pair $(f,g)\in \mathcal{Q}\times\mathcal{Q}$, the sum is still an internal operation because it is the sum of two non decreasing functions, while the product between a scalar $k$ and a {\em qf} is internal (i.e. returns a {\em qf}) only if $k\geq 0$. Using these operators, and considering the sum of quantile functions as the $F$ in eq. \ref{eq:CHIM} , it is possible to affirm that $\bar{\Phi}(t)$ (associated with $M_W(y)$) is the {\em Chisini} mean of a set of quantile functions which is invariant with respect to the sum of {\em qf}s.\\
Being $M_W(y)$ a density function, we may derive its basic statistics like the mean and the variance as follows:
\begin{equation}\label{mean of mean}
    \mu_{\bar y}=\int\limits_{-\infty}^{+\infty}{y\cdot \bar \phi(y)dy}=\int\limits_{0}^{1}{\bar \Phi^{-1}(t)dt}=\frac{1}{n}\sum\limits_{i=1}^n \int\limits_{0}^{1}{\Phi_i^{-1}(t)dt}=\frac{1}{n}\sum\limits_{i=1}^n\mu_i,
    \end{equation}
which correspond to the mean of the means of the distribution. The variance of $M_W(y)$ is formulated as follows:
\begin{equation}\label{var of mean}
\begin{array}{l}
  \sigma^2_{\bar y}=\int\limits_{-\infty}^{+\infty}{y^2\cdot \bar \phi(y)dy}-\left(\mu_{\bar y}\right)^2=\int\limits_{0}^{1}{ \left[\bar \Phi^{-1}(t)\right]^2dt}-\left(\int\limits_{0}^{1}{\bar \Phi^{-1}(t)dt}\right)^2= \\
   = \int\limits_0^1 {{{\left[ {\frac{1}{n}\sum\limits_{i = 1}^n {\Phi _i^{ - 1}(t)} } \right]}^2}dt}  - {\left( {\frac{1}{n}\sum\limits_{i = 1}^n {\mathop \smallint \limits_0^1 \Phi _i^{ - 1}(t)dt} } \right)^2}=\\ = \frac{1}{{{n^2}}}{ {\sum\limits_{i = 1}^n {\sum\limits_{j = 1}^n {\int\limits_0^1\Phi _i^{ - 1}(t)\Phi _j^{ - 1}(t)} } } dt}  - {\left( {\frac{1}{n}\sum\limits_{i = 1}^n {{\mu _i}} } \right)^2}.
\end{array}
    \end{equation}
    For simplifying the last formula, we present the following formulation of the product of two quantile functions.
\begin{rmk} \label{def_prod} Given two quantile functions $\Phi_i^{ - 1}(t)$ and $\Phi_j^{ - 1}(t)$, associated with two pdf's $\phi_i(y)$ and $\phi_i(y)$ with means $\mu_i$ and $\mu_j$ and standard deviations $\sigma_i$  and $\sigma_j$, the product is defined as follows:
\begin{equation}\label{eq:inner}
    \langle\Phi_i^{ - 1},\Phi_j^{ - 1}\rangle=\int\limits_0^1 { \Phi_i^{ - 1}(t) \Phi_j^{ - 1}(t) dt}=\rho _{i,j}\sigma_i\sigma_j+\mu_i\mu_j.
\end{equation}
\end{rmk}
The proof is straightforward using algebra from eq. (\ref{eq:rhoQQ}).\\
Using this result, we obtain a final formulation of the variance of $M_W(y)$  as follows:
\begin{eqnarray}\label{eq:sigma2_MW}
\sigma^2_{\bar y}&=&\frac{1}{n^2}\sum\limits_{i = 1}^n {\sum\limits_{j = 1}^n {\left[ {{\rho _{i,j}}{\sigma _i}{\sigma _j}} \right]} }= \nonumber \\
&=&\sum\limits_{i=1}^n{\left[\frac{\sigma_i}{n}\right]^2}+\frac{2}{n^2}\sum\limits_{i = 1}^{n-1} {\sum\limits_{j >i} {\left[ {{\rho _{i,j}}{\sigma _i}{\sigma _j}} \right]} }.
\end{eqnarray}
It is worth noting that if all the distributions have the same shape then  $\rho _{i,j}=1$ for each couple of distributions and the variance of $M_W$ reaches its maximum value. The minimum value is clearly obtained when all the observed data are points (i.e. $\sigma_i=0$ for each $i=1,\ldots,n$), thus:
\begin{equation}\label{var of mean2}
0\leq \sigma^2_{\bar y} \leq\left(\frac{1}{n}\sum\limits_{i=1}^n \sigma_i\right)^2.
\end{equation}
We do not investigate the computation of the further moments of the distribution associated with $M_W(y)$ because it requires further considerations beyond the topic of this paper.
However, it is worth noting that $M_W(y)$ is a distribution having a shape similar to all the $n$ distributions: if we have single-valued data (points), $M_W(y)$ is a point (i.e. it generalizes the arithmetic mean of a set of standard data), if we have interval-valued data, $M_W(y)$ is an interval-valued description, if we have histogram-valued data, $M_W(y)$ is a histogram.

\paragraph{The variance of Y with respect to $M_W(y)$.}
 Given $M_W(y)$, the mean of a set of $n$ units described by the distributional symbolic variable $Y$, we define the variance of $Y$  as the mean of the squared Wasserstein distance between each distribution $y(i)$ and $M_W(y)$. In this sense, the variance of $Y$ corresponds to the {\em Frech\'{e}t} criterion in eq. (\ref{eq:FREM}) with $w_i=\frac{1}{n}$. Using the definition of product between two {\em qf}s in definition \ref{def_prod}, we denote with $S_W^2(Y)$ the variance of $Y$ which is computed as follows:
\begin{equation}\label{eq:variance}
\begin{array}{l}
S_W^2(Y)=\frac{1}{n}\sum\limits_{i=1}^{n}{d_W^2(\phi_i(y),M_W(y))}=\\
=\frac{1}{n}\sum_{i=1}^{n}\int\limits_0^1 {\left( {\Phi_i^{-1}(t)  -\bar \Phi^{-1}(t) } \right)^2dt}
=SM_W^2(Y)+SV_W^2(Y)=\\
=\left[ {\frac{1}{n}\sum\limits_{i = 1}^n {\mu _i^2}  - {{\left( {{\mu _{\bar y}}} \right)}^2}} \right] + \left[ {\frac{1}{n}\sum\limits_{i = 1}^n {\sigma _i^2}  - \frac{1}{{{n^2}}}\sum\limits_{i = 1}^n {\sum\limits_{j = 1}^n {{\rho _{i,j}}{\sigma _i}{\sigma _j}} } } \right].
\end{array}
\end{equation}
We note that $S_W^2(Y)$ is the sum of two positive independent sources of variability:
\begin{description}
  \item[$SM_W^2(Y)={\left[\frac{1}{n}\sum\limits_{i = 1}^n {\mu _i^2}  - {{\left( {{\mu _{\bar y}}} \right)}^2}\right]}$] is the variance of the means of the $n$ distributions;
  \item[$SV_W^2(Y)={\left[\frac{1}{n}\sum\limits_{i = 1}^n {\sigma _i^2}  - \frac{1}{{{n^2}}}\sum\limits_{i = 1}^n {\sum\limits_{j = 1}^n {{\rho _{i,j}}{\sigma _i}{\sigma _j}} } \right]}$] is a measure of variance related to the (squared) differences of the internal variability of the $n$ distributions. $SV_W^2(Y)$ is always positive. Considering that ${\sum\limits_{j = 1}^n {{\rho _{i,j}}{\sigma _i}{\sigma _j}} }$ is maximum when, for each couple of distributions, $\rho_{i,j}=1$, (i.e.,  all the distributions have the same shape), and if $\sigma_i>0$ $i=1,\ldots,n$, we observe that the minimum value of $SV_W^2(Y)$ is:
      $$
      SV_W^2(Y)={\frac{1}{n}\sum\limits_{i = 1}^n {\sigma _i^2}  - \left[\frac{1}{{{n}}}\sum\limits_{i = 1}^n  {\sigma _i} \right]^2 },
      $$
      that is the variance of the standard deviations of the $n$ distributions. Finally, $SV_W^2(Y)$  is equal to zero in two cases: when all the distributions are identically distributed except for their means or when all the data are single valued.
\end{description}
Differently from  $S^2(Y)$ in eq. \ref{var_MIX}, $S_W^2(Y)$ is also equal to zero when all the $n$ distributions are identical and positive $\sigma_i$'s. Secondly, comparing $S^2$ in eq. (\ref{var_Bi}) (or its simplified version of eq. (\ref{var_MIX})) with  $S_W^2(Y)$, it is clear that $S^2(Y)$ depends only from the means and the standard deviations of the compared distributions, while $S_W^2(Y)$ depends also from the different shapes of the compared distributions. Finally, $S^2(Y)$ is generally greater than $S_W^2(Y)$. In fact, rewriting $S^2$ we observe that:
$${S^2(Y)} = \left[ {\frac{1}{n}\sum\limits_{i = 1}^n {\mu _i^2}  - {{\left( {{\mu _{\bar y}}} \right)}^2}} \right] + \frac{1}{n}\sum\limits_{i = 1}^n { {\sigma _i^2} }  \ge S_W^2(Y)$$
being the difference between the two indices equal to:
$${S^2(Y)} - S_W^2(Y)=\frac{1}{{{n^2}}}\sum\limits_{i = 1}^n {\sum\limits_{j = 1}^n {{\rho _{i,j}}{\sigma _i}{\sigma _j}} }.$$
The difference depends from the shape and the standard deviations of the distributions.\\
\paragraph{The standard deviation} According to the $d_W$, a generalization of the standard deviation of the numerical modal multi-valued variable $Y$ observed for a set of $n$ units  is the following:
\begin{equation}\label{st_dev_W}
  S_W(Y)=\sqrt{\left[ {\frac{1}{n}\sum\limits_{i = 1}^n {\mu _i^2}  - {{\left( {{\mu _{\bar y}}} \right)}^2}} \right] + \left[ {\frac{1}{n}\sum\limits_{i = 1}^n {\sigma _i^2}  - \frac{1}{{{n^2}}}\sum\limits_{i = 1}^n {\sum\limits_{j = 1}^n {{\rho _{i,j}}{\sigma _i}{\sigma _j}} } } \right].}
\end{equation}
Using the sum and the product by a scalar for quantile functions, it is straightforward to show that, given the variable $Y$, its standard deviation $S_W(Y)$ respects the following properties:
\begin{enumerate}
  \item Positivity: $S_W(Y) \geq0$.
  \item If all data are identically distributed (i.e. have the same modal multi-valued numerical description) then $$S_W(Y)=0.$$
  \item Given two real numbers $h\geq 0$ and $k$ and being $Z=h\cdot Y +k$ a transformation of the $Y$ variable, the corresponding $S_{W}(Z)$ standard deviation is: $$S_{W}(Z)=h\cdot S_W(Y).$$
\end{enumerate}

The two novel basic statistics $M_W(y)$ and $S_W^2(Y)$ reach the objective of better considering the double source of the variability of a symbolic variable $Y$ observed on a set of $n$ units: the mean {\em internal to the data} variability is expressed by the variability of the $M_W(y)$ distribution, while the $S_W^2(Y)$ takes into consideration only the differences among the distributions, i.e. it measure the {\em between} variability.\\
From a computational point of view, the difficulties of computing an exact value for $S_W^2(Y)$ is related to the possibility of computing the $\rho_{i,j}$. \citet{IrVe2006, IrLeVe2006} proposed a closed form for computing the squared Wasserstein distance between two histogram-valued data and from that formation it is also possible to derive the closed form related to $\rho_{i,j}$. The computation is also done in a time that is linear with respect to the number of bins of the histograms. The same is possible for interval-valued data, considering them as data described by {\em trivial} histograms (i.e., histograms having only one-bin).  For data described by different types of density functions the $\rho_{i,j}$ can be derived analytically only when all the {\em qf}s can be expressed in closed forms (for example, this is not possible for the Normal distribution). In all the other cases, numerical methods can be applied for approximating $\rho(i,j)$.\\
Another result presented in \cite{IrVe2006, IrLeVe2006} is related to the variance decomposition in a framework of clustering analysis of histogram-valued data. \cite{IrVe2006, IrLeVe2006}, after showing that the $\ell_2$ Wasserstein distance is an extension of the Euclidean distance between quantile functions, the authors showed that it was possible to obtain a decomposition of the variability of a set of histograms according to the Huygens theorem of decomposition of the inertia and used such properties for extending some clustering methods for standard data to histogram-valued data.

\section{Measures of  interdependence}\label{sec:interdep}
In this section, we consider how to extend the classic measure of association between two single real valued variables, like the covariance and the correlation indices, to a couple of numeric modal variables.\\
Starting from the \citet{BerGoup2000} approach, \citet{BilDid2006} proposed a formulation of the covariance between interval or histogram valued variables. For example, let $Y_1$ and $Y_2$ be two interval-valued variables, such that the generic $l$-th unit ($l=1,\ldots,n$) is described by the ordered pair of descriptions $y(i)=\{y_1(i),y_2(i)\}$ where $y_1(i)=[a_{i1};b_{i1}]$ and $y_2(i)=[a_{i2};b_{i2}]$ are intervals, the $C_B(Y_1,Y_2)$ covariance index has the following formulation:
\begin{equation}\label{covar_bil_did}
    C_B(Y_1,Y_2)=\frac{1}{3n}\sum_{i=1}^{n}{G_{i1}G_{i2}\sqrt{Q_{i1}Q_{i2}}}
\end{equation}
where
\begin{equation}\label{cond_covar}
\begin{array}{l}
 Q_{i1}  = \left( {a_{i1}  - \bar Y _1 } \right)^2  + \left( {a_{i1}  - \bar Y _1 } \right)\left( {b_{i1}  - \bar Y _1 } \right) + \left( {b_{i1}  - \bar Y _1 } \right)^2  \\
 G_{i1}  = \left\{ {\begin{array}{*{20}c}
   { - 1} & {if} & {\frac{a_{i1}+b_{i1}}{2}  \le \bar Y _1 }  \\
   1 & {if} & {\frac{a_{i1}+b_{i1}}{2}  > \bar Y _1 }  \\
\end{array}} \right. \\
Q_{i2}  = \left( {a_{i2}  - \bar Y _2 } \right)^2  + \left( {a_{i2}  - \bar Y _1 } \right)\left( {b_{i2}  - \bar Y _2 } \right) + \left( {b_{i2}  - \bar Y _2 } \right)^2  \\
 G_{i2}  = \left\{ {\begin{array}{*{20}c}
   { - 1} & {if} & {\frac{a_{i2}+b_{i2}}{2}  \le \bar Y _2 }  \\
   1 & {if} & {\frac{a_{i2}+b_{i2}}{2}  > \bar Y _2 }  \\
\end{array}} \right. \\
 \end{array}
\end{equation}
and $\bar Y _1$ and $\bar Y _2$ are the means calculated according to eq. (\ref{mu_Bil}). The authors \cite[Eq. 4.19, pag. 136]{BilDid2006}  extended $C_B(Y_1,Y_2)$ to data described by a couple of histogram valued variables, considering them as weighted combination of intervals. Like for the univariate statistics, such statistics can be  brought back to an approach based on mixture of bivariate distributions. Further, the proposed measures do not consider clearly the different sources of variability of a set of multi-valued symbolic data. \Citet{BilDid2006} also propose a measure of correlation that is computed as follows:
\begin{equation}\label{corr_bil_did}
    R_B(Y_1,Y_2)=\frac{C_B(Y_1,Y_2)}{S(Y_1)S(Y_2)}.
\end{equation}

\paragraph{The covariance based on $\ell_2$ Wasserstein metric}
Using the $\ell_2$ Wasserstein metric and the associated product of {\em qf}s defined in Eqn. (\ref{eq:inner}), we propose an alternative approach for the measure of the covariance and of the correlation between two symbolic variables and for solving some of the above mentioned deficiencies of the \cite{BilDid2006} approach. Let $Y_1$ and $Y_2$ be two modal numeric variables describing a set of $n$ units, the generic $i-th$ unit is described by the ordered pair ${y(i)=\{y_1(i),y_2(i)\}}$ where $y_1(i)=\phi_i(y_1)$ and $y_2(i)=\phi_i(y_2)$ are density functions, with respective means equal to $\mu_{i1}$ and $\mu_{i2}$, and standard deviations $\sigma_{i1}$ and $\sigma_{i2}$. With each $\phi_i(y_1)$ (resp. $\phi_i(y_2)$) is associated the corresponding {\em cdf} $\Phi_i(y_1)$ (resp. $\Phi_i(y_2)$) and the respective {\em qf} denoted $\Phi^{-1}_{i1}(t)$ (resp.$\Phi^{-1}_{i2}(t)$).\\
 We denote with $C_W(Y_1,Y_2)$ the empirical covariance between $Y_1$ and $Y_2$ based on the $\ell_2$ Wasserstein metric as follows:
\begin{equation}\label{eq:covarW}
 C_W(Y_1,Y_2)= \frac{1}{n}\sum\limits_{i = 1}^n {\int\limits_0^1 {\left[{ \Phi_{i1}^{ - 1} \left( t \right) - \bar \Phi_1^{ - 1} \left( t \right)} \right]\cdot\left[ {\Phi_{i2}^{ - 1} \left( t \right) - \bar \Phi_{2}^{ - 1} \left( t \right)} \right]dt} }
\end{equation}
where $\bar \Phi_{1}^{ - 1}( t )$ (resp. $\bar \Phi_{2}^{ - 1}( t )$) is the {\em qf} associated to the {\em Fr\'{e}chet} mean distribution based on the $\ell_2$ Wasserstein metric $M_W(y_1)$ (resp. $M_W(y_2)$).  Given the $i$-th and the $j$-th generic unit, we rewrite the indices of $\rho_{(\cdot,\cdot)}$ in Eq. (\ref{eq:rhoQQ}) such that $\rho_{i1,j2}$ denotes the correlation of the {\em qf}s $\Phi^{-1}_{i1}(t)$ and $\Phi^{-1}_{j2}(t)$, while $\rho_{\cdot1,\cdot2}$ denotes the correlation of the {\em qf}s associated with $M_W(y_1)$ (i.e., $\bar\Phi^{-1}_{i1}(t)$) and $M_W(y_2)$ (i.e., $\Phi^{-1}_{j2}(t)$). Using the proposed notation and the product of two {\em qf}s defined in Eq. (\ref{eq:inner}), $C_W(Y_1,Y_2)$ can be rewritten as follows:
\begin{equation}\label{eq:finalCOV}
    \begin{array}{l}
C_W(Y_1,Y_2)=CM_W(Y_1,Y_2)+CV_W(Y_1,Y_2)=\\
\left( {\frac{1}{n}\sum\limits_{i = 1}^n {{\mu _{i1}}{\mu _{i2}}}  - {\mu _{{{\bar y}_1}}}{\mu _{{{\bar y}_2}}}} \right) +\\
+ \left( {\frac{1}{n}\sum\limits_{i = 1}^n {{\rho _{i1,i2}}{\sigma _{i1}}{\sigma _{i2}} - \frac{1}{{{n^2}}}\sum\limits_{i = 1}^n {\sum\limits_{j = 1}^n {{\rho _{i1,j2}}} } } {\sigma _{i1}}{\sigma _{j2}}} \right).
  \end{array}
\end{equation}
Similarly to the variance in Eq. (\ref{eq:covarW}), we see that the index $C_W(Y_1,Y_2)$ is the sum of two kinds of covariance: the first denoted with $CM_W(Y_1,Y_2)$ is  clearly the covariance of the means, while  the latter, denoted with $CV_W(Y_1,Y_2)$, is related to a sort of covariance in variability. In this case, it is possible that the two components have different signs but, on the other hand, it allows one to better consider different aspects for the comparison of multi-valued modal variables. For example, if all the distributions have the same shape (they are all normally distributed) then all the $\rho$'s are equal to $1$ and $C_W(Y_1,Y_2)$ can be simplified as the sum of the covariance of the means plus the covariance of the standard deviations, as follows:
\begin{equation}\label{eq:addend}
    \begin{array}{l}
C_W(Y_1,Y_2)=\left( {\frac{1}{n}\sum\limits_{i = 1}^n {{\mu _{i1}}{\mu _{i2}}}  - {\mu _{{{\bar y}_1}}}{\mu _{{{\bar y}_2}}}} \right)
+ \left( {\frac{1}{n}\sum\limits_{i = 1}^n {\sigma _{i1}}{\sigma _{i2}} - \sigma_{\bar y_1}\sigma_{\bar y_2} } \right).
  \end{array}
\end{equation}
If all the distributions, as well as being identical in shape, also have the same standard deviation, then the second term becomes equal to zero. As in the classical case, it is noteworthy that if the covariance $C_W(Y_1,Y_1)$ is calculated on the same variable it coincide with the variance $S_W^2(Y_1)$ of the variable. In a different formulation, $C_W(Y_1,Y_2)$ has been used in \citep{VeIR08} for proposing a Mahalanobis-Wasserstein distance for clustering data described by histogram variables, and showing its analogies with the Mahalanobis distance for standard variables.

\paragraph{The correlation index} The last index we present is an extension of the correlation index for modal numerical variable. Given the covariance $C_W(Y_1,Y_2)$ between two numeric modal variables $Y_1$ and $Y_2$ and the respective standard deviations $S_W(Y_1)$ and $S_W(Y_2)$, we denote the correlation index with $R_W(Y_1,Y_2)$ and is calculated as follows:
\begin{equation}\label{eq:correlW}
    R_W(Y_1,Y_2)=\frac{C_W(Y_1,Y_2)}{{S_W(Y_1)\cdot S_W(Y_2)}}.
\end{equation}
$R_W(Y_1,Y_2)$ depends from $C_W(Y_1,Y_2)$, but the ratio does not allow one to separate completely the contribution of the means and of the variability of the distributions. However, we propose to consider $R_W(Y_1,Y_2)$ as the sum of two parts: the first, denoted with $RM_W(Y_1,Y_2)$, is mostly related to the means, while the latter, denoted with $RV_W(Y_1,Y_2)$, is mostly related to the variability of the distributions as follows:
\begin{eqnarray}\label{eq:correlW2}
    R_W(Y_1,Y_2)=RM_W(Y_1,Y_2)&+&RV_W(Y_1,Y_2)= \nonumber\\
    =\frac{CM_W(Y_1,Y_2)}{{S_W(Y_1)\cdot S_W(Y_2)}}&+&\frac{CV_W(Y_1,Y_2)}{{S_W(Y_1)\cdot S_W(Y_2)}}.
\end{eqnarray}
Finally, it is noteworthy to show that $R_W(Y_1,Y_2)=1$ if, and only if, $Y_1$ is equal to $Y_2$, while this is not generally true for the correlation measure $R_B(Y_1,Y_2)$ in eq. (\ref{corr_bil_did}).

\section{Example}\label{sec:apply}
\begin{figure}
\begin{centering}
\includegraphics[width=0.95\textwidth]{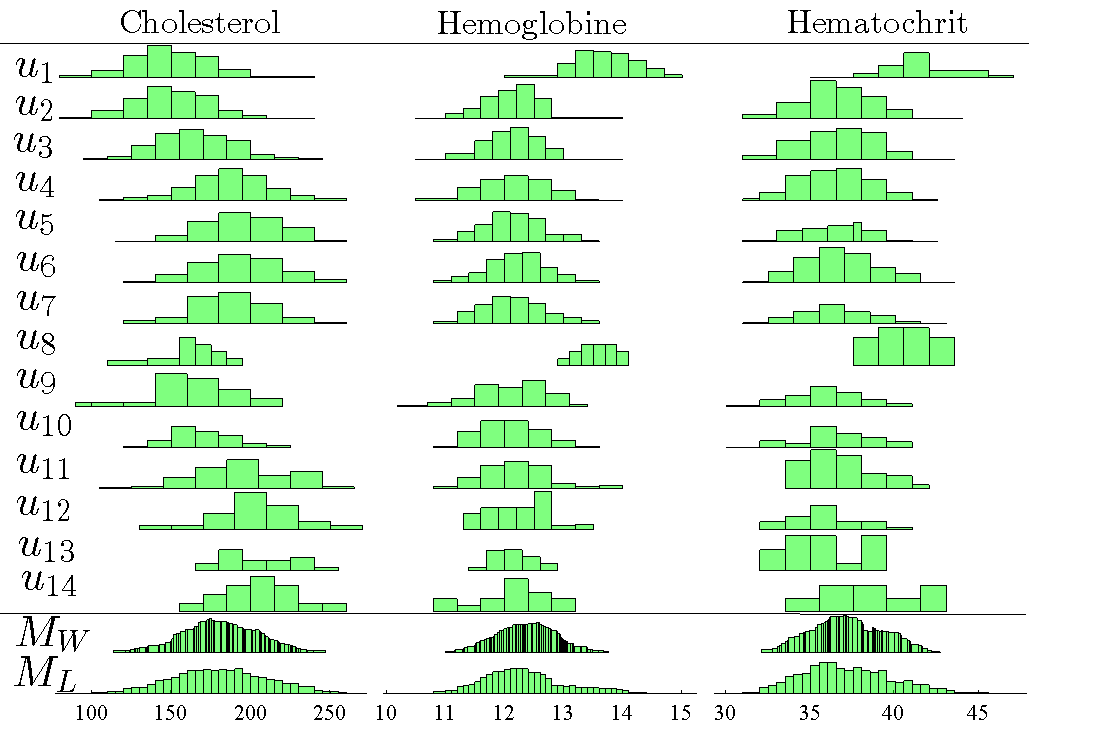}
\end{centering}
\caption{Blood dataset: histogram representation of the data. The last two rows  represents the $M_W$ and the $M_L$ mean histograms.}\label{Fig: Blood_data}
\end{figure}

The proposed univariate and bivariate basic statistics give new interpretative tools for the analysis of numeric multi-valued data. For showing it, we compute the statistics using a data set presented in \cite{BilDid2006}\footnote{We joined together the three tables describing the three histogram variables that are presented in different sections of the book.}, and we compare the obtained results with the basic statistics therein proposed. We call this dataset as \emph{Blood dataset}.\\
\begin{table}[htbp]
  \centering
  \caption{Blood dataset: Cholesterol}
  {\footnotesize
  \resizebox{\textwidth}{!}{
    \begin{tabular}{p{1.7cm}p{10.5cm}}
    \hline
           Unit \cr {Gender-Age} &\multicolumn{1}{c} {Frequency histogram}\\
    \hline
    $u_1$:  F-20   & [80, 100), 0.025; [100, 120), 0.075; [120, 135), 0.175; [135, 150), 0.250; [150, 165), 0.200;  [165, 180), 0.162; [180, 200), 0.088; [200, 240], 0.025 \\
    $u_2$: F-30   &\{[80, 100), 0.013; [100, 120), 0.088; [120, 135), 0.154; [135, 150), 0.253; [150, 165), 0.210; [165, 180), 0.177; [180, 195), 0.066; [195, 210), 0.026; [210, 240], 0.013\} \\
    $u_3$: F-40   & \{[95, 110), 0.012; [110, 125), 0.029; [125, 140), 0.113; [140, 155), 0.206; [155, 170), 0.235; [170, 185), 0.186; [185, 200), 0.148; [200, 215), 0.043; [215, 230), 0.020; [230, 245], 0.008\} \\
    $u_4$: F-50   & {[105, 120), 0.009; [120, 135), 0.026; [135, 150), 0.046; [150, 165), 0.105; [165, 180), 0.199; [180, 195), 0.248; [195, 210), 0.199; [210, 225), 0.100; [225, 240), 0.045; [240, 260], 0.023} \\
    $u_5$: F-60   & {[115, 140), 0.012; [140, 160), 0.069; [160, 180), 0.206; [180, 200), 0.300; [200, 220), 0.255; [220, 240), 0.146; [240, 260], 0.012} \\
    $u_6$:  F-70   & {[120, 140), 0.017; [140, 160), 0.083; [160, 180), 0.206; [180, 200), 0.294; [200, 220), 0.250; [220, 240), 0.117; [240, 260], 0.033} \\
    $u_7$: F-80+   & {[120, 140), 0.036; [140, 160), 0.065; [160, 180), 0.284; [180, 200), 0.325; [200, 220), 0.213; [220, 240), 0.065; [240, 260], 0.012} \\
    $u_8$: M-20   & {[110, 135), 0.143; [135, 155), 0.143; [155, 165), 0.286; [165, 175), 0.214; [175, 185), 0.143; [185, 195], 0.071} \\
    $u_9$: M-30   & {[90, 100), 0.022; [100, 120), 0.044; [120, 140), 0.044; [140, 160), 0.333; [160, 180), 0.289; [180, 200), 0.179; [200, 220], 0.089} \\
    $u_{10}$: M-40   & {[120, 135), 0.018; [135, 150), 0.109; [150, 165), 0.327; [165, 180), 0.255; [180, 195), 0.182; [195, 210), 0.073; [210, 225], 0.036} \\
    $u_{11}$: M-50   & {[105, 125), 0.019; [125, 145), 0.020; [145, 165), 0.118; [165, 185), 0.216; [185, 205), 0.294; [205, 225), 0.137; [225, 245], 0.176; [245, 265], 0.020} \\
    $u_{12}$: M-60   & {[130, 150), 0.041; [150, 170), 0.042; [170, 190), 0.167; [190, 210), 0.375; [210, 230), 0.250; [230, 250), 0.083; [250, 270], 0.042} \\
    $u_{13}$: M-70   & {[165, 180), 0.105; [180, 195), 0.316; [195, 210), 0.158; [210, 225), 0.158; [225, 240), 0.210; [240, 255], 0.053} \\
    $u_{14}$: M-80+   & {[155, 170), 0.067; [170, 185), 0.133; [185, 200), 0.200; [200, 215), 0.267; [215, 230), 0.200; [230, 245), 0.067; [245, 260], 0.066} \\
    \hline
    \end{tabular}}
    }
  \label{tab:Chol}
\end{table}
\begin{table}[htbp]
  \centering
  \caption{Blood dataset: Hemoglobin}
  {\footnotesize
  \resizebox{\textwidth}{!}{
     \begin{tabular}{p{1.7cm}p{10.5cm}}
    \hline
           Unit \cr {Gender-Age} &\multicolumn{1}{c} {Frequency histogram}\\
    \hline
   $u_1$:  F-20   &  {[12.0, 12.9), 0.050; [12.9, 13.2), 0.112; [13.2, 13.5), 0.212; [13.5, 13.8), 0.201; [13.8, 14.1), 0.188; [14.1, 14.4), 0.137; [14.4, 14.7), 0.075; [14.7, 15.0], 0.025}\\
 $u_2$: F-30   &{[10.5, 11.0), 0.007; [11.0, 11.3), 0.039; [11.3, 11.6), 0.082; [11.6, 11.9), 0.174; [11.9, 12.2), 0.216; [12.2, 12.5), 0.266; [12.5, 12.8), 0.157; [12.8, 14.0], 0.059}\\
$u_3$: F-40   &{[10.5, 11.0), 0.009; [11.0, 11.5), 0.084; [11.5, 11.8), 0.148; [11.8, 12.1), 0.217; [12.1, 12.4), 0.252; [12.4, 12.7), 0.180; [12.7, 13.0), 0.087; [13.0, 14.0], 0.023}\\
 $u_4$: F-50   &{[10.5, 11.2), 0.046; [11.2, 11.6), 0.134; [11.6, 12.0), 0.222; [12.0, 12.4), 0.259; [12.4, 12.8), 0.219; [12.8, 13.2), 0.105; [13.2, 13.6), 0.012; [13.6, 14.0], 0.003}\\
 $u_5$: F-60   &{[10.8, 11.2), 0.028; [11.2, 11.5), 0.081; [11.5, 11.8), 0.133; [11.8, 12.1), 0.231; [12.1, 12.4), 0.219; [12.4, 12.7), 0.182; [12.7, 13.0), 0.061; [13.0, 13.3), 0.057; [13.3, 13.6], 0.008}\\
 $u_6$:  F-70   &{[10.8, 11.1), 0.022; [11.1, 11.4), 0.050; [11.4, 11.7), 0.078; [11.7, 12.0), 0.183; [12.0, 12.3), 0.228; [12.3, 12.6), 0.233; [12.6, 12.9), 0.117; [12.9, 13.2), 0.067; [13.2, 13.6], 0.022}\\
 $u_7$: F-80+   &{[10.8, 11.2), 0.029; [11.2, 11.5), 0.095; [11.5, 11.8), 0.148; [11.8, 12.1), 0.213; [12.1, 12.4), 0.207; [12.4, 12.7), 0.160; [12.7, 13.0), 0.077; [13.0, 13.3), 0.047; [13.3, 13.6], 0.024}\\
 $u_8$: M-20   &{[12.9, 13.1), 0.071; [13.1, 13.3), 0.143; [13.3, 13.5), 0.214; [13.5, 13.7), 0.217; [13.7, 13.9), 0.212; [13.9, 14.1], 0.143}\\
 $u_9$: M-30   &{[10.2, 10.7), 0.022; [10.7, 11.1), 0.045; [11.1, 11.5), 0.089; [11.5, 11.9), 0.222; [11.9, 12.3), 0.200; [12.3, 12.7), 0.267; [12.7, 13.1), 0.133; [13.1, 13.4], 0.022}\\
$u_{10}$: M-40   &{[10.8, 11.2), 0.018; [11.2, 11.6), 0.163; [11.6, 12.0), 0.273; [12.0, 12.4), 0.273; [12.4, 12.8), 0.182; [12.8, 13.2), 0.073; [13.2, 13.6], 0.018}\\
$u_{11}$: M-50   &{[10.8, 11.2), 0.020; [11.2, 11.6), 0.118; [11.6, 12.0), 0.235; [12.0, 12.4), 0.275; [12.4, 12.8), 0.235; [12.8, 13.2), 0.059; [13.2, 13.6), 0.020; [13.6, 14.0], 0.038}\\
$u_{12}$: M-60   &{[11.3, 11.6), 0.125; [11.6, 11.9), 0.166; [11.9, 12.2), 0.166; [12.2, 12.5), 0.167; [12.5, 12.8), 0.292; [12.8, 13.2), 0.042; [13.2, 13.5], 0.042}\\
$u_{13}$: M-70   &{[11.4, 11.7), 0.053; [11.7, 12.0), 0.315; [12.0, 12.3), 0.316; [12.3, 12.6), 0.211; [12.6, 12.9], 0.105}\\
 $u_{14}$: M-80+   &{[10.8, 11.2), 0.133; [11.2, 11.6), 0.067; [11.6, 12.0), 0.134; [12.0, 12.4), 0.333; [12.4, 12.8), 0.200; [12.8, 13.2], 0.133}\\
    \hline
    \end{tabular}}
    }
  \label{tab:Hem}
\end{table}
\begin{table}[htbp]
  \centering
  \caption{Blood dataset: Hematocrit}
  {\footnotesize
  \resizebox{\textwidth}{!}{
    \begin{tabular}{p{1.7cm}p{10.5cm}}
    \hline
           Unit \cr {Gender-Age} &\multicolumn{1}{c} {Frequency histogram}\\
    \hline
   $u_1$:  F-20   & {[35.0, 37.5), 0.025; [37.5, 39.0), 0.075; [39.0, 40.5), 0.188; [40.5, 42.0), 0.387; [42.0, 45.5), 0.287; [45.5, 47.0), 0.038}\\
 $u_2$: F-30   &{[31.0, 33.0), 0.046; [33.0, 35.0), 0.171; [35.0, 36.5), 0.295; [36.5, 38.0), 0.243; [38.0, 39.5), 0.170; [39.5, 41.0), 0.072; [41.0, 44.0], 0.003}\\
$u_3$: F-40   &{[31.0, 33.0), 0.049; [33.0, 35.0), 0.203; [35.0, 36.5), 0.223; [36.5, 38.0), 0.241; [38.0, 39.5), 0.209; [39.5, 41.0), 0.069; [41.0, 43.5], 0.006}\\
$u_4$: F-50   &{[31.0, 32.0), 0.011; [32.0, 33.5), 0.066; [33.5, 35.0), 0.194; [35.0, 36.5), 0.231; [36.5, 38.0), 0.248; [38.0, 39.5), 0.168; [39.5, 41.0), 0.068; [41.0, 42.5], 0.014}\\
$u_5$: F-60   &{[31.0, 33.0), 0.037; [33.0, 34.5), 0.178; [34.5, 36.0), 0.215; [36.0, 37.5), 0.247; [37.5, 38.0), 0.101; [38.0, 39.5), 0.182; [39.5, 41.0), 0.028; [41.0, 42.5], 0.012}\\
 $u_6$:  F-70   &{[31.0, 32.5), 0.011; [32.5, 34.0), 0.089; [34.0, 35.5), 0.200; [35.5, 37.0); 0.272; [37.0, 38.5), 0.228; [38.5, 40.0), 0.122; [40.0, 41.5), 0.072; [41.5, 43.5], 0.006}\\
$u_7$: F-80+   &{[31.0, 32.5), 0.018; [32.5, 34.0), 0.107; [34.0, 35.5), 0.195; [35.5, 37.0), 0.302; [37.0, 38.5), 0.189; [38.5, 40.0), 0.136; [40.0, 41.5), 0.041; [41.5, 43.0], 0.012}\\
 $u_8$: M-20   &{[37.5, 39.0), 0.214; [39.0, 40.5), 0.286; [40.5, 42.0), 0.286; [42.0, 43.5], 0.214}\\
 $u_9$: M-30   &{[30.0, 32.0), 0.022; [32.0, 33.5), 0.111; [33.5, 35.0), 0.178; [35.0, 36.5), 0.311; [36.5, 38.0), 0.222; [38.0, 39.5), 0.111; [39.5, 41.0], 0.045}\\
$u_{10}$: M-40   &{[30.0, 32.0), 0.018; [32.0, 33.5), 0.109; [33.5, 35.0), 0.073; [35.0, 36.5), 0.327; [36.5, 38.0), 0.218; [38.0, 39.5), 0.164; [39.5, 41.0], 0.091}\\
$u_{11}$: M-50   &{[33.5, 35.0), 0.215; [35.0, 36.5), 0.294; [36.5, 38.0), 0.255; [38.0, 39.5), 0.118; [39.5, 41.0), 0.098; [41.0, 42.0], 0.020}\\
$u_{12}$: M-60   &{[32.0, 33.5), 0.125; [33.5, 35.0), 0.208; [35.0, 36.5), 0.375; [36.5, 38.0), 0.125; [38.0, 39.5), 0.125; [39.5, 41.0], 0.042}\\
$u_{13}$: M-70   &{[32.0, 33.5), 0.158; [33.5, 35.0), 0.263; [35.0, 36.5), 0.263; [36.5, 38.0), 0.053; [38.0, 39.5], 0.263}\\
$u_{14}$: M-80+   &{[33.5, 35.5), 0.133; [35.5, 37.5), 0.267; [37.5, 39.5), 0.267; [39.5, 41.5), 0.133; [41.5, 43.0], 0.200}\\
    \hline
    \end{tabular}
    }}
  \label{tab:Hema}
\end{table}
The dataset consists of histogram-valued observations for three histogram variables: \emph{Cholesterol} (in Table \ref{tab:Chol}), \emph{Hemoglobin} (in Table \ref{tab:Hem}), and \emph{Hematocrit} (in table \ref{tab:Hema}). Each observation represents one of 14 $gender\times age$ groups of patients, where gender is denoted with $M=Male$ and $F=Female$ and the age group refers to decades from 20 to 80 (and over) years. As reported in \cite{BilDid2006}, the histogram values resulted from aggregating a portion of a very large dataset containing (classical) values for individuals living in a certain region of the USA. Finally, it is worth noting that only univariate histograms are available.\\
Tables \ref{tab:Chol}, \ref{tab:Hem} and \ref{tab:Hema} could not be easy to read because of the multi-valued nature of the data. To facilitate a better view of the dataset,  in Fig. \ref{Fig: Blood_data} we have represented graphically the histograms related to the 14 observations for the three variables\footnote{We remark that each histogram has a mass equal to 1 and the heights of the histograms are only graphically scaled for the presentation of data.}. Further, for a rapid comparison, the last two rows show the histogram representation of the two proposed {\em Fr\'{e}chet} means: $M_W$ and $M_L$.
For example, observing Fig. \ref{Fig: Blood_data} we  note that for the \emph{Cholesterol} variable and for the \emph{Female} typologies of patients (from $u_1$ to $u_7$), the distributions are similar in shape and  symmetric, and that the means increase from the youngest typology ($u_1$) to the older one ($u_8$). The distributions of the  \emph{Male} typologies are more skewed than the \emph{Female} ones, and also have different shapes.
With regard to the \emph{Hemoglobin} and \emph{Hematocrit} variables, it is possible to note that the younger typologies ($u_1$ and $u_8$) hold distributions with means different from other types.
Looking at the mean values, we see that $M_W(y)$ tends to best summarize the average shape of the distributions, while $M_L(y)$ suffers from the fact that it represents a finite mixture of distributions. Table \ref{tab:Bstat_data} shows the main basic statistics of the observed histograms and of the two means, and we observe that the basic statistics of $M_W(t)$ are closer to the average basic statistics (the number in bold) of the distributions than the basic statistics of $M_L(y)$.\\
\begin{table}
  \centering
  \caption{Blood dataset: basic statistics of the symbolic data. We used the third standardized moment as the skewness index and the fourth standardized moment minus 3 for the kurtosis one. The average values of the indices are in bold.}
  \resizebox{\textwidth}{!}{
    \begin{tabular}{|l|rrr|rrr|rrr|rrr|}
    \hline
          & \multicolumn{ 3}{|c|}{Mean} & \multicolumn{ 3}{|c|}{Standard deviation} & \multicolumn{ 3}{|c|}{Skewness} & \multicolumn{ 3}{|c|}{Kurtosis} \\
          & Chol. & Hemo. & Hemat. & Chol. & Hemo. & Hemat. & Chol. & Hemo. & Hemat. & Chol. & Hemo. & Hemat. \\
          \hline
    u1    & 150.10 & 13.695 & 41.526 & 26.34 & 0.550 & 2.197 & 0.229 & -0.209 & 0.119 & 0.367 & 0.080 & -0.001 \\
    u2    & 150.71 & 12.158 & 36.497 & 25.28 & 0.528 & 2.122 & 0.197 & 0.388 & -0.075 & 0.230 & 1.097 & -0.279 \\
    u3    & 164.96 & 12.134 & 36.549 & 25.33 & 0.507 & 2.230 & 0.163 & 0.061 & -0.131 & -0.047 & 0.603 & -0.502 \\
    u4    & 186.51 & 12.133 & 36.480 & 26.66 & 0.585 & 2.198 & -0.100 & -0.168 & 0.072 & 0.183 & -0.193 & -0.491 \\
    u5    & 194.03 & 12.145 & 36.341 & 25.21 & 0.520 & 2.098 & -0.150 & 0.058 & -0.014 & -0.352 & -0.225 & -0.417 \\
    u6    & 193.20 & 12.205 & 36.703 & 26.56 & 0.523 & 2.182 & -0.027 & -0.114 & 0.165 & -0.319 & -0.094 & -0.393 \\
    u7    & 187.14 & 12.141 & 36.504 & 24.59 & 0.552 & 2.191 & -0.056 & 0.185 & 0.183 & 0.044 & -0.282 & -0.272 \\
    u8    & 159.62 & 13.557 & 40.500 & 19.84 & 0.300 & 1.636 & -0.662 & -0.153 & 0.000 & -0.150 & -0.859 & -1.043 \\
    u9    & 164.43 & 12.088 & 35.914 & 26.49 & 0.622 & 2.114 & -0.305 & -0.439 & -0.028 & 0.158 & -0.168 & -0.274 \\
    u10   & 170.06 & 12.092 & 36.457 & 20.01 & 0.527 & 2.248 & 0.374 & 0.241 & -0.208 & -0.170 & -0.386 & -0.343 \\
    u11   & 194.22 & 12.214 & 36.720 & 30.16 & 0.597 & 2.002 & -0.167 & 0.524 & 0.496 & -0.264 & 0.445 & -0.511 \\
    u12   & 203.36 & 12.245 & 35.815 & 26.22 & 0.509 & 2.008 & -0.174 & 0.081 & 0.346 & 0.369 & -0.661 & -0.373 \\
    u13   & 205.67 & 12.150 & 35.750 & 22.50 & 0.334 & 2.165 & 0.217 & 0.226 & 0.210 & -1.083 & -0.634 & -1.142 \\
    u14   & 205.48 & 12.120 & 38.450 & 23.54 & 0.616 & 2.616 & 0.106 & -0.440 & 0.090 & -0.437 & -0.584 & -1.056 \\
    Aver. & \textbf{180.68} & \textbf{12.363} & \textbf{37.157} & \textbf{24.91} & \textbf{0.519} & \textbf{2.143} & \textbf{-0.025} & \textbf{0.017} & \textbf{0.082} & \textbf{-0.105} & \textbf{-0.133} & \textbf{-0.507} \\
          &       &       &       &       &       &       &       &       &       &       &       &  \\
    $M_W$ & 180.68 & 12.363 & 37.157 & 24.78 & 0.516 & 2.133 & -0.025 & -0.008 & 0.082 & -0.210 & -0.290 & -0.602 \\
    $M_L$ & 180.68 & 12.363 & 37.157 & 31.66 & 0.739 & 2.743 & -0.036 & 0.568 & 0.409 & -0.250 & 0.232 & -0.095 \\
    \hline
    \end{tabular}
    }
  \label{tab:Bstat_data}
\end{table}
In Table \ref{tab:unvSTAT}, we reported the univariate statistics as proposed by \cite{BerGoup2000} \cite{BilDid2006} and as proposed in this paper.
As described in section \ref{sec:Univariate_stats}, although the $S_2(Y)$ variance can be decomposed into the variance of the means and the mean of the variances, it does not allow us to understand the different sources of variability of the symbolic variable. The basic statistics based on the $\ell_2$ Wasserstein distance allow one to observe that the average $M_W(y)$ can be considered as that observation with a description (a distribution) which is \emph{intermediate} with respect the others.
\begin{table}
  \centering
  \caption{Blood dataset: basic univariate statistics for each variable.}
  {\begin{tabular}{lrrr}
    \hline
          & \multicolumn{3}{c}{Billard}   \\
        Indices  & Cholesterol & Hemoglobin & Hematocrit \\
          \hline
        \multicolumn{4}{c}{Sample means}  \\
    $\bar Y$  & 180.68 & 12.363 & 37.157 \\
            \multicolumn{4}{c}{Variability statistics}  \\
    $S^2$ & 1002.34 & 0.5466 & 7.526 \\
    Variance of means & 374.864 & 0.2686 & 2.893 \\
    \% of Variance      & 37.40\% & 49.14\% & 38.44\% \\
    Mean of variances & 627.476 & 0.278 & 4.633 \\
    \% of Variance      & 62.60\% & 50.86\% & 61.56\% \\
    $S$  & 31.658 & 0.7393 & 2.7434 \\
    \hline
          &       &       &  \\
          \hline
          & \multicolumn{3}{c}{Wasserstein}  \\
   Indices    & Cholesterol & Hemoglobin & Hematocrit \\
      \hline
           \multicolumn{4}{c}{Main statistics of $M_W(y)$}  \\

    $\mu_{\bar y}$ & 180.68 & 12.363 & 37.157 \\
    $\sigma_{\bar y}$ & 24.78 & 0.516 & 2.133 \\
         \multicolumn{4}{c}{Variability statistics} \\
    $S^2_W(Y)$ & 388.138 & 0.2802 & 2.978 \\
    $SM^2_W(Y)$& 374.864 & 0.2686 & 2.893 \\
    \% of $S^2_W(Y)$& 96.58\% & 95.86\% & 97.15\% \\
    $SV^2_W(Y)$& 13.274 & 0.0116 & 0.0849 \\
    \% of $S^2_W(Y)$& 3.42\% & 4.14\% & 2.85\% \\
          &       &       &  \\
    $S_W(Y)$ & 19.701 & 0.5294 & 1.7257 \\
    \hline
    \end{tabular}}
  \label{tab:unvSTAT}
\end{table}
The $S^2_W(Y)$ statistics of variability (together with the variability statistics of $M_W(y)$) allow us to better interpret the different source of variability of a set of symbolic data. Firstly, we note that $S^2_W(Y)$ is always lower than $S_2(Y)$  because a portion of the variability present in $S_2(Y)$ is incorporated by the variability of $M_W(y)$. Observing the composition of $S^2_W(Y)$, we note that for each variable the percentage of variance due to the means is always higher than 95\%. In particular, the \emph{Hematocrit} variable is the one that presents the lower variability component as regards the variability of the distributions of the 14 units ($SV^2_W(Y)$ is the 2.85\% of $S^2_W(Y)$). Indeed, looking at the histograms observed for  the \emph{Hematocrit} variable there is less difference between the distributions in terms of internal variability, for example, with respect to the histograms of the \emph{Cholesterol} variable, where the distributions from $u_8$ to $u_14$ are very different in shape both among them and with respect to distributions from $u_1$ to $u_7$ (that, among them, are very similar in shape and internal variability).\\
Tables \ref{tab:Biv_Bill_stat}, \ref{tab:CovarianceW} and \ref{tab:CorrelW} show the bivariate statistics discussed in this article and calculated for the Blood dataset. In Table \ref{tab:Biv_Bill_stat} we reported the covariance and correlation calculated according to the methodology proposed by \citet{BilDid2006}.
\begin{table}
  \centering
  \caption{Blood dataset: Billard \cite{BilDid2006} bivariate statistics}
  \resizebox{\textwidth}{!}{
    \begin{tabular}{lrrrclrrr}
           &  &  &  \\
           \multicolumn{4}{c}{Covariance}& & \multicolumn{4}{c}{Correlation}\\
       $C_B(Y_i,Y_j)$   & Chol. & Hemo. & Hemat. & &$R_B(Y_i,Y_j)$ & Chol. & Hemo. & Hemat. \\
       \cline{1-4} \cline{6-9}

    Chol. & 1002.339 & -5.266 & -15.6928  & &Chol. & 1     & -0.225 & -0.1807 \\
       Hemo. &       & 0.547 & 0.8205  & &Hemo. &       & 1     & 0.4045 \\
    Hemat. &       &       & 7.5265 & &Hemat. &       &       & 1 \\
               \cline{1-4} \cline{6-9}
    \end{tabular}}
  \label{tab:Biv_Bill_stat}
\end{table}
Looking at the correlations, it seems there are no significant correlations between all the pairs of variables. However, in hematology it is known that there should be a direct link between \emph{Hematocrit} and \emph{Hemoglobin}. In fact, the hematocrit test determines how much of the total blood volume contains red blood cells. The red blood cells are basically vessels for hemoglobin, so there is a very direct relationship between hemoglobin and hematocrit.
\begin{table}
  \centering
  \caption{Blood dataset: Wasserstein based covariance statistics}

    \begin{tabular}{lrrr}
          & \multicolumn{ 3}{c}{} \\
      $C_W(Y_i,Y_j)$    & Cholesterol & Hemoglobin & Hematocrit  \\
    \hline
    Cholesterol & 388.138 & -5.001 & -14.920  \\
    Hemoglobin &       & 0.280 & 0.826  \\
    Hematocrit &       &       & 2.978 \\
    \hline
    &       &       &  \\
    \end{tabular}
    \resizebox{\textwidth}{!}{
    \begin{tabular}{lrrrclrrr}
    $CM_W(Y_i,Y_j)$     & Chol. & Hemo. & Hemat. & & $CV_W(Y_i,Y_j)$     &Chol. & Hemo. & Hemat.\\
\cline{1-4} \cline{6-9}
    Chol. & 374.864 & -5.179 & -15.086 & &Chol. & 13.274 & 0.178 & 0.165\\
    Hemo. &       & 0.269 & 0.813 & &Hemo. &       & 0.012 & 0.014 \\
    Hemat. &       &       & 2.893  & &Hemat. &       &       & 0.085\\
  \cline{1-4} \cline{6-9}
    \end{tabular}}
  \label{tab:CovarianceW}
\end{table}
\begin{table}
  \centering
  \caption{Blood dataset: Wasserstein based correlation statistics}
    \begin{tabular}{lrrr}
          & \multicolumn{ 3}{c}{} \\
       $R_W(Y_i,Y_j)$   & Cholesterol & Hemoglobin & Hematocrit \\
          \hline
    Cholesterol & 1     & -0.4795 & -0.4389 \\
    Hemoglobin &       & 1     & 0.9049 \\
    Hematocrit &       &       & 1 \\
    \hline
     &       &       &  \\
     \end{tabular}
    \resizebox{\textwidth}{!}{
    \begin{tabular}{lrrrclrrr}
    $RM_W(Y_i,Y_j)$     & Chol. & Hemo. & Hemat. & & $RV_W(Y_i,Y_j)$     &Chol. & Hemo. & Hemat.\\
\cline{1-4} \cline{6-9}
    Chol. & 0.9658 & -0.4966 & -0.4437 & &Chol. & 0.0342 & 0.0171 & 0.0049\\
    Hemo. &       & 0.9585 & 0.8896 & &Hemo. &       & 0.0415 & 0.0153 \\
    Hemat. &       &       & 0.9715  & &Hemat. &       &       & 0.0285\\
  \cline{1-4} \cline{6-9}
    \end{tabular}}
  \label{tab:CorrelW}
\end{table}
Unfortunately, the results in Table \ref{tab:Biv_Bill_stat} do not confirm this relationship ($R_B(Hemo.,Hemat) = 0.4045$). Otherwise, using the measures of the covariance (Table  \ref{tab:CovarianceW} ) and of the correlation (Table \ref{tab:CorrelW}) based on the $\ell_2$ Wasserstein metric, the relationship between \emph{Hematocrit} and \emph{Hemoglobin} returns more evident ($R_W(Hemo.,Hemat)  = 0.9049$).
Observing the decomposition of the correlation in Table \ref{tab:CorrelW}, we can affirm that the correlation between the correlation is given in good part by the correlation component due to the means ($RM_W(Hemo.,Hemat)  = 0.8896$) and a part also due to the internal variability of the distributions ($RV_W(Hemo.,Hemat)  = 0.0153$).

\section{Conclusions and future research}\label{sec:concl}
In this paper, we have presented new basic univariate and bivariate statistics for numerical modal variable. We discussed the possibility of also treating the other numeric symbolic variable as a numeric modal variable in order to extend the basic statistics to all the numeric symbolic variables. The basic statistics are based on the $\ell_2$ Wasserstein distance between univariate distributions. The new statistics are compared with those proposed by \cite{BerGoup2000} and \citet{BilDid2006} and showing, using an application on a medical dataset, their interpretative properties, and emphasising the capacity of the novel statistics of taking into account the different source of variability of a multivariate symbolic dataset.\\
The bivariate statistics need a deeper reflection. In the literature, a clear and univocal definition of the relationship between two symbolic variables is still missing, thus also the indices for measuring it are still in embryonal methodological phase. Finally, we consider that a deeper study about the inferential methodologies based on such kind of data can give a great impulse to the research.
\appendix
\section{Proof of the decomposition of the $\ell_2$ squared Wasserstein distance.}\label{Appendix}
Let $\phi_i(y)$ and $\phi_{i'}(y)$ be two density functions having finite the first two moments. The $\phi_i(y)$ density function is in a one-to-one correspondence with the cumulative distribution function $\Phi_i(y)$ and the quantile function $\Phi_i^{-1}(t)$ (the inverse of the distribution function). The expected value of $\phi_i(y)$ is denoted with $\mu_i$ and the standard deviations with  $\sigma_i$.
In this appendix we prove that:
\begin{equation}\label{eq:proof}
\begin{array}{l}
d_W^2 (\phi_i(y) ,\phi_{i'}(y) )
= \int\limits_0^1 {\left[ {\Phi_i^{ - 1} (t) - \Phi_{i'}^{ - 1} (t)} \right]^2 dt}=\\
=\underbrace {\left( {\mu _i - \mu _{i'} }
\right)^2}_{Location} + \underbrace{\underbrace {\left( {\sigma _i - \sigma _{i'}
} \right)^2}_{Size} + \underbrace {2 \sigma_i \sigma_{i'} (1 -
\rho_{i,i'} )}_{Shape}}_{Variability}.
\end{array}
\end{equation}
First of all we note that
\begin{equation}\label{eq:mu}
{\mu _i} = \int\limits_{ - \infty }^{ + \infty } {y\cdot{\phi _i}(y)dy}  = \int\limits_{ - \infty }^{ + \infty } {yd{\Phi _i}(y)}  = \int\limits_{ - \infty }^{ + \infty } {\Phi _i^{ - 1}({\Phi _i}(y))d{\Phi _i}(y)}  = \int\limits_0^1 {\Phi _i^{ - 1}(t)dt},
\end{equation}
where $t = \Phi(y)$, $\Phi(-\infty)=0$, $\Phi(+\infty)=1$  and considering that $ y = \Phi^{ - 1} (\Phi(y)) = \Phi^{ - 1} (t)$.
Analogously, for $\sigma^2$ we have that:
  \begin{equation}\label{eq:ss}
  {\sigma _i}^2 = \int\limits_{ - \infty }^{ + \infty } {{y^2}{\phi _i}(y)dy - \mu _i^2}  = \int\limits_{ - \infty }^{ + \infty } {{{\left[ {\Phi _i^{ - 1}({\Phi _i}(y))} \right]}^2}d{\Phi _i}(y)}  - \mu _i^2 = \int\limits_0^1 {{{\left[ {\Phi _i^{ - 1}(t)} \right]}^2}dt - \mu _i^2} .\end{equation}
We develop the squared term of the distance, and using eqs. (\ref{eq:mu}) and (\ref{eq:ss})  we obtain:
\begin{equation}\label{eq:DEV1}
\begin{array}{l}
d_W^2\left( {{\phi _i}(y),{\phi _{i'}}(y)} \right) = \int\limits_0^1 {{{\left[ {\Phi _i^{ - 1}(t) - \Phi _{i'}^{ - 1}(t)} \right]}^2}dt}  = \\
 = \int\limits_0^1 {{{\left[ {\Phi _i^{ - 1}(t)} \right]}^2}dt}  + \int\limits_0^1 {{{\left[ {\Phi _{i'}^{ - 1}(t)} \right]}^2}dt}  - 2\int\limits_0^1 {\Phi _i^{ - 1}(t)\cdot\Phi _{i'}^{ - 1}(t)dt}  = \\
 = \sigma _i^2 + \mu _i^2 + \sigma _{i'}^2 + \mu _{i'}^2 - 2\int\limits_0^1 {\Phi _i^{ - 1}(t)\cdot\Phi _{i'}^{ - 1}(t)dt}
\end{array}
\end{equation}

Now we introduce the following quantity:

\begin{equation}\label{AppEq5}
\begin{array}{c}
{\rho _{i,i'}} = \frac{{\int\limits_0^1 {\left( {\Phi _i^{ - 1}(t) - {\mu _i}} \right)\left( {\Phi _{i'}^{ - 1}(t) - {\mu _{i'}}} \right)dt} }}{{\sqrt {\left[ {\int\limits_0^1 {{{\left( {\Phi _i^{ - 1}(t) - {\mu _i}} \right)}^2}dt} } \right]\left[ {\int\limits_0^1 {{{\left( {\Phi _{i'}^{ - 1}(t) - {\mu _{i'}}} \right)}^2}dt} } \right]} }} = \\=\frac{{\int\limits_0^1 {\left( {\Phi _i^{ - 1}(t) - {\mu _i}} \right)\left( {\Phi _{i'}^{ - 1}(t) - {\mu _{i'}}} \right)dt} }}{{{\sigma _i}{\sigma _{i'}}}}
 = \int\limits_0^1 {\frac{{\left( {\Phi _i^{ - 1}(t) - {\mu _i}} \right)}}{{{\sigma _i}}}\frac{{\left( {\Phi _{i'}^{ - 1}(t) - {\mu _{i'}}} \right)}}{{{\sigma _{i'}}}}dt}  =\\= \frac{{\int\limits_0^1 {\Phi _i^{ - 1}(t)\Phi _{i'}^{ - 1}(t)dt}  - {\mu _i}{\mu _{i'}}}}{{{\sigma _i}{\sigma _{i'}}}}
\end{array}
\end{equation}

that is the correlation of two series of data where each couple of observations is represented respectively by the $t-th$ quantile of the first distribution and the $t-th$ quantile of the second. In this sense we may consider it as the correlation between quantile functions represented by the curve of the infinite quantile points in a QQ plot.
It is worth noting that, if $\sigma_i$ and $\sigma_{i'}$ are positive,  $0 < \rho _{i,i'}  \le 1$ and is equal to 1 when the two standardized series of quantiles are the same, or, in other words, when the two distributions are identical except for the means and the standard deviations (i.e., they are two uniforms, two normal distributions, etc.).

Using the last term of $\rho_{i,i'}$ in eq. (\ref{AppEq5}), we observe that
$$
{\int\limits_0^1 {\Phi _i^{ - 1}(t)\cdot\Phi _{i'}^{ - 1}(t)dt} }=\rho_{i,i'}\sigma _i\sigma _{i'} + {\mu _i}{\mu _{i'}}.
$$
Thus, we continue developing eq.(\ref{eq:DEV1}) as follows
 	
\begin{equation}\label{DEV2}
\begin{array}{l}
d_W^2\left( {{\phi _i}(y),{\phi _{i'}}(y)} \right) = \sigma _i^2 + \mu _i^2 + \sigma _{i'}^2 + \mu _{i'}^2 - 2\int\limits_0^1 {\Phi _i^{ - 1}(t)\cdot\Phi _{i'}^{ - 1}(t)dt}  = \\
 = \sigma _i^2 + \mu _i^2 + \sigma _{i'}^2 + \mu _{i'}^2 - 2\left[ {{\rho _{i,i'}}{\sigma _i}{\sigma _{i'}} + {\mu _i}{\mu _{i'}}} \right] = \\
 = \left( {\mu _i^2 + \mu _{i'}^2 - 2{\mu _i}{\mu _{i'}}} \right) + \sigma _i^2 + \sigma _{i'}^2 - 2{\rho _{i,i'}}{\sigma _i}{\sigma _{i'}}.
\end{array}
\end{equation}
Finally, adding and subtracting  $2\sigma_i \sigma_{i'}$ we prove eq. (\ref{eq:proof})
\begin{equation}\label{AppEq7}
\begin{array}{l}
d_W^2\left( {{\phi _i}(y),{\phi _{i'}}(y)} \right) = \\
 = \left( {\mu _i^2 + \mu _{i'}^2 - 2{\mu _i}{\mu _{i'}}} \right) + \sigma _i^2 + \sigma _{i'}^2 - 2{\rho _{i,i'}}{\sigma _i}{\sigma _{i'}} + 2{\sigma _i}{\sigma _{i'}} - 2{\sigma _i}{\sigma _{i'}} = \\
 = {\left( {{\mu _i} - {\mu _{i'}}} \right)^2} + \left( {\sigma _i^2 + \sigma _{i'}^2 - 2{\sigma _i}{\sigma _{i'}}} \right) + 2{\sigma _i}{\sigma _{i'}} - 2{\rho _{i,i'}}{\sigma _i}{\sigma _{i'}} = \\
 = {\left( {{\mu _i} - {\mu _{i'}}} \right)^2} + {\left( {{\sigma _i} - {\sigma _{i'}}} \right)^2} + 2{\sigma _i}{\sigma _{i'}}\left( {1 - {\rho _{i,i'}}} \right).\qed
\end{array}
\end{equation}

\bibliographystyle{plain}

\end{document}